\documentclass[12pt]{iopart}

\usepackage{iopams}  
\usepackage{graphicx}
\usepackage{xcolor}
\usepackage{upgreek}
\begin{document}

\title[Stochastic energetics of a colloidal particle trapped in a viscoelastic bath]{Stochastic energetics of a colloidal particle trapped in a viscoelastic bath}

\author{Farshad Darabi$^{1}$, Brandon R. Ferrer$^{1}$, and Juan Ruben Gomez-Solano$^{1,*}$}

\address{$^{1}$Instituto de F\'isica, Universidad Nacional Aut\'onoma de M\'exico, Ciudad de M\'exico, C\'odigo Postal 04510, Mexico,}

\ead{$^*$ r\_gomez@fisica.unam.mx}
\vspace{10pt}
\begin{indented}
\item[]September 2023
\end{indented}

\begin{abstract}
We investigate the statistics of the fluctuations of the energy transfer between an overdamped Brownian particle, whose motion is confined by a stationary harmonic potential, and a surrounding viscoelastic fluid at constant temperature. We derive an analytical expression for the probability density function of the energy exchanged with the fluid over a finite time interval, which implicitly involves the friction memory kernel that encodes the coupling with such a non-Markovian environment, and reduces to the well known expression for the heat distribution in a viscous fluid. We show that, while the odd moments of this distribution are zero, the even moments can be explicitly expressed in terms of the autocorrelation function of the particle position, which generally exhibits a non-mono-exponential decay when the fluid bath is viscoelastic. Our results are verified by experimental measurements for an optically-trapped colloidal bead in semidilute micellar and polymer solutions, finding and excellent agreement for all time intervals over which the energy exchange takes place. 
\end{abstract}

%
\vspace{2pc}

%
%
%

\section{Introduction}

Understanding energy exchanges in micron- and sub-micron-sized systems subject to fluctuations, e.g. colloidal particles, living cells, biopolymers, molecular motors, and small electronic circuits, is of prime importance in many disciplines of natural and applied sciences \cite{ciliberto2017}. 
In the last couple of decades, a significant progress has been made towards achieving this objective  thanks to the advent of stochastic thermodynamics, which extends macroscopic concepts such as heat, work and entropy production to the level of single stochastic realizations of such processes~\cite{ciliberto2017,sekimoto1998,seifert2012}. On this basis, fundamental constraints for the fluctuations of these thermodynamic quantities~\cite{jarzynski1997,crooks1998,hatano2001,evans2002,seifert2005}, as well as fluctuation-dissipation relations around out-of-equilibrium states~\cite{harada2005,chetrite2008,baiesi2009,seifert2010,verley2011,altaner2016}, have been derived and as such, they represent refinements of the classical principles of thermodynamics. These theoretical relations have been tested in various experiments in small systems~\cite{wang2002,liphardt2002,carberry2004,blickle2006,toyabe2007,gomezsolano2009,gomezsolano2012}, which confirmed their validity under rather general non-equilibrium conditions. More recently, stochastic thermodynamics has also proved extremely helpful in studying energy fluxes and entropy production in more complex settings, such as in active matter~\cite{ganguly2013, speck2016,dabelow2019,szamel2019,cao2022}, and systems with anomalous diffusion \cite{hartich2021,khadem2022}, thus providing a strong  theoretical framework to describe small-scale thermodynamic processes arbitrarily away from equilibrium.

For mesoscopic systems in contact with reservoirs at constant temperature, e.  g. colloids, vesicles, and macromolecules in aqueous solution,
energy is continuously transferred between the system and the bath due to thermal collisions with the surrounding molecules even in the absence of external driving forces~\cite{fogedby2020}. Then, in many situations it is not enough to only know the second-law bounds imposed on the energy flows or the entropy production by the fluctuations relations~\cite{jarzynski1997,crooks1998,hatano2001,evans2002,seifert2005} but also the detailed shape of their corresponding probability distributions. This is of special interest for processes taking place during short time intervals over which fluctuations are expected to largely exceed the mean values of the thermodynamic quantity of interest. Along these lines, a number of investigations using stochastic thermodynamics have been carried out over the past years in order to have a grasp of the statistics of the energy exchanged as heat between a Brownian system and its environment in absence of applied work. For instance, Fokker-Planck equations for the probability density function of the heat have been derived in presence of arbitrary confining potentials, whose long-time asymptotic solutions were experimentally verified for colloidal particles trapped in water by optical tweezers~\cite{imparato2007}. Analytical expressions for the probability distribution of the heat transferred during an arbitrary time interval for the same system have also been obtained using path integrals \cite{chatterjee2010,chatterjee2011}. Moreover, the heat probability distribution has been determined for overdamped Brownian particles in non-stationary states relaxing toward thermal equilibrium after a temperature quench \cite{gomezsolano2011,crisanti2017}. In addition, the statistics of the steady heat fluxes for systems in contact with two thermostats at different temperatures have been investigated for quantum harmonic oscillators \cite{denzler2018,chen2021}, RC electric circuits~\cite{ciliberto_2013,ciliberto2013}, pairs of hydrodynamically-coupled colloidal particles~\cite{berut2016}, and harmonic networks~\cite{saito2011,kundu2011,fogedby2012,dhar2015}. Other effects on the heat distribution for Brownian particles, e.g. nonlinear potentials~\cite{fogedby2009,paraguassu_2021,paraguassu_2022}, inertia~\cite{rosinberg2016,salazar2016,paraguassu2022,colmenares2022,paraguassu_2023}, relativistic motion~\cite{paraguassu2021}, and non-isothermal transformations~\cite{paraguassu2023}, have also been addressed theoretically. Furthermore, heat fluctuations have  been studied for active matter systems, such as active chains in viscous heat baths~\cite{gupta2021}, Brownian particles embedded in active media~\cite{goswami2019,goswami2022}, and activity-driven harmonic chains~\cite{sarkar2023}.

It is worth mentioning that in most of the investigations on the heat probability distribution for thermostatted Brownian systems, there is a clear separation between the time scales of the system and those of the bath, which results in an effective Markovian description of the slow degrees of freedom of the system, for which the derivation of the heat distribution is straightforward. Nevertheless, the Markovian property is not possessed by a broad diversity of soft matter systems ranging from dense colloidal suspensions to polymeric fluids, which exhibit slow relaxations due to their crowded macromolecular microstructure~{\cite{larson1999}}. This leads to memory effects on the dynamics of Brownian-particle systems dispersed in such media, which are commonly described at equilibrium by the generalized Langevin equation, where the environment acts as a non-Markovian bath~\cite{zwanzig1973}. Surprisingly, to the best of our knowledge, there is only a single theoretical work where the calculation of the first two moments of the distribution of the energy exchanged between a Brownian particle in a harmonic trap and its viscoelastic-fluid surroundings is carried out based on the generalized Langevin equation~\cite{chatterjee2009}. This leaves open the question of what is the full shape of the probability distribution under such non-Markovian conditions for an arbitrary time interval over which the stochastic energy transfer takes place.  
We point out that finding such a probability distribution for a Brownian system in equilibrium with a single non-Markovian bath is a first step towards the detailed knowledge on the statistics of the energy exchange in more intricate non-equilibrium situations, e.g. under time-dependent driving forces, reservoirs at different temperatures, and gradients. In turn, such problems are of great relevance in many applications at mesoscopic scales, \emph{e.g}, the micromanipulation of colloidal probes in soft materials~\cite{gomezsolano2015,roichman2021}, the operation of Brownian heat engines in complex media~\cite{krishnamurthy2016,gomezsolano2021,guevara2023,nalupurackal2023}, and the controlled microswimming in viscoelastic environments \cite{gomezsolano2016,narinder2018,saad2019,raman2023}.

In view of the above considerations, the main purpose of this paper is to get insights into the effect of the non-Markovianity of a heat bath on the stochastic energetics for small thermostatted systems. To this end, we focus on a model system, namely, a micron-sized bead trapped by stationary optical tweezers in a viscoelastic fluid kept at constant temperature, whose motion is governed by the generalized Langevin equation. {Based on the characteristic functional of the stochastic thermal force acting on the particle}, we are able to derive an analytical expression in the overdamped limit for the energy transferred between the particle and the surrounding fluid over an arbitrary time interval. Our theoretical results are verified by experimental measurements in viscoelastic fluids such as aqueous worm-like micellar and polymer solutions. 

The paper is organized as follows. In~\sref{sect:model} we present the model for the motion of a Brownian particle subject to a harmonic potential in a viscoelastic fluid, from which we find explicit formulae for some statistical quantities that describe the stochastic dynamics at thermal equillibrium. Then, in~\sref{sec:pdf} we derive an analytical expressions for the probability density function of the energy transferred from the fluid to the particle, and discuss its connection with previously obtained expressions for similar systems in the case of a Markovian heat bath. In~\sref{sec:exp} we describe the experimental setup that we use to analyze the stochastic energetics of a colloidal bead harmonically trapped in some complex fluids with different viscoelastic responses, and make a comparison with our analytical results. Finally, in~\sref{sect:conclude}, we summarize our main results and make some further physical remarks.

\section{Model}\label{sect:model}
We consider a spherical Brownian particle of mass $m$ and radius $a$, which is embedded in an incompressible viscoelastic fluid medium with stress relaxation modulus $G(t)$ and mass density $\rho$ at constant temperature $T$. The viscoelastic fluid might be composed of macromolecules, e.g. polymer chains, micelles, or colloidal nanoparticles, dispersed in a viscous solvent, whose relaxation modulus is generally a slowly decaying function of $t$ such that $G(t \rightarrow \infty) = 0$, i.e. it behaves as a liquid in the long-time limit~{\cite{larson1999}}. We focus on the motion of a single coordinate of the center of mass of the particle, $x$, which is subject to a stationary harmonic potential of constant stiffness $\kappa$, i.e. $U(x) = \frac{1}{2}\kappa x^2$, as commonly implemented in experiments by means of optical tweezers. At thermal equilibrium, $x$ evolves stochastically in time according to the generalized Langevin equation \cite{zwanzig1973,indei2012}
\begin{equation}\label{eq:GLE}
    m_{\mathrm{eff}}\ddot{x}(t) = -\int_0^t dt' \, \Gamma(t-t') \dot{x}(t') -\partial_x U(x)|_{x = x(t)} + \zeta(t).
\end{equation}
The left-hand side of~\eref{eq:GLE} corresponds to the inertial force on the particle, where $x(t)$ is its instantaneous position at time $t\ge 0$ starting from the initial conditions $x_0 = x(0)$ and $v_0 = \dot{x}(0)$ at time $t = 0$. This inertial term involves the effective mass $m_{\mathrm{eff}} = m + \frac{2}{3}\rho \pi a^3$ that includes, in addition to the particle mass $m$, half of the mass of the fluid displaced by the particle~{\cite{kim2005}}.  Besides, the first term on the right-hand side of~\eref{eq:GLE} represents the coarse-grained drag force exerted on the trapped particle at time $t$ by all the surrounding fluid particles. Here, $\Gamma(t-t')$ is a memory kernel that encodes the delayed effect of the fluid at times $0 \le t' \le t$, with $\Gamma(t - t') = 0$ for $t' > t$ by causality. Furthermore, $  -\partial_x U(x)|_{x = x(t)} = -\kappa x(t)$ corresponds to the value of the conservative force at time $t$ that derives from the harmonic potential $U(x)$. Moreover, $\zeta(t)$ is a Gaussian colored noise that accounts for the thermal collisions of the fluid particles, whose mean and autocorrelation function satisfy at thermal equilibrium
\begin{eqnarray}\label{eq:FD2nd}
    \langle \zeta(t) \rangle & = & 0, \nonumber\\
    \langle \zeta(t) \zeta(t') \rangle & = & k_B T \Gamma(|t-t'|),
\end{eqnarray}
respectively. The Laplace transform of the memory kernel, $\tilde{\Gamma}(s) = \int_0^{\infty} dt\, e^{-st} \Gamma(t)$ is related to the fluid viscoelasticity by the expression~\cite{procopio2023}
\begin{equation}\label{eq:GSLap}
    \tilde{\Gamma}(s) = 6\pi a \tilde{\eta}(s) + 6\pi a^2 \sqrt{\rho s \tilde{\eta}(s)}.
\end{equation}
The first term on the right-hand side of~\eref{eq:GSLap} corresponds to conventional Stokes law for the drag force on a sphere moving in a fluid with a viscosity $\tilde{\eta}(s)$ that is dependent on the Laplace frequency, $s$, which is determined by the Laplace transform of the fluid's relaxation modulus, $\tilde{\eta}(s) \equiv \tilde{G}(s)  = \int_0^{\infty} dt \, e^{-st} G(t)$, whereas the second term represents the Basset-Boussinesq force that originates from the motion of the displaced fluid~\cite{landau1959,zwanzig1970}. 

{
We focus on the overdamped limit of~\eref{eq:GLE}, i.e. $m_{\mathrm{eff}}, \rho \rightarrow 0$, which is a very good approximation in many colloidal experiments where $a \sim 10^{-6}$~m, in such a way that the accessible frequencies to detect the single-particle motion are commonly $|s| \ll |\tilde{\Gamma}(s)|/m_{\mathrm{eff}} \sim  10^6 \,\mathrm{rad \, s}^{-1}$ and $|s| \ll |\tilde{\eta}(s)|/(\rho a^2) \sim  10^6 \,\mathrm{rad \, s}^{-1}$. In such a case, inertial terms can be totally neglected in the particle dynamics and, consequently, Eq.~(\ref{eq:GLE}) becomes
\begin{equation}\label{eq:ODGLE}
	0 = -\int_0^t dt' \, \Gamma(t-t') \dot{x}(t') - \kappa x(t) + \zeta(t),
\end{equation}
where the Laplace transform of the memory kernel $\Gamma(t)$ can be approximated as
\begin{equation}
    \tilde{\Gamma}(s)  =  6\pi a \tilde{\eta}(s). \label{eq:ODGamma}
\end{equation}
Note that in the case of a Newtonian fluid of frequency-independent viscosity $\tilde{\eta} (s) = \eta$, the relaxation modulus is $G(t) = 2\eta \delta(t)$, where $\delta(t)$ is the Dirac delta function. This yields the well-known overdamped Langevin equation for a Brownian particle trapped in a viscous medium
\begin{equation}\label{eq:Lang}
    \gamma \dot{x}(t) = -\kappa x(t) + \zeta(t),
\end{equation}
where $\gamma = 6\pi a \eta$ is the friction coefficient and  $\zeta(t)$ reduces to a zero-mean delta-correlated Gaussian noise, i.e. $\langle \zeta(t) \zeta(t')\rangle = 2k_B T \gamma \delta(t-t')$.

By taking the Laplace transform of each term in~\eref{eq:ODGLE}, for the initial condition $x(0) = x_0$ the solution of the Laplace transform of $x(t)$, i.e. $\tilde{x}(s) = \int_0^{\infty} dt\, e^{-st} x(t)$, reads
\begin{equation}\label{eq:Lapx}
    \tilde{x}(s) = \tilde{\chi}(s) x_0  + \tilde{\varsigma}(s) \tilde{\zeta}(s),
\end{equation}
where the functions $\tilde{\chi}(s)$ and $\varsigma(s)$ are defined in Laplace domain by the expressions
\begin{equation}\label{eq:chi}
    \tilde{\chi}(s) = \frac{\tilde{\Gamma}(s)}{ s \tilde{\Gamma}(s) + \kappa},
\end{equation}
and
\begin{equation}\label{eq:sigma}
    \tilde{\varsigma}(s) = \frac{1}{s\tilde{\Gamma}(s) + \kappa},
\end{equation}
respectively. Therefore, the solution in time domain, $x(t)$, for a given stochastic realization of the thermal noise $\zeta(t')$ during the interval $0 \le t' \le t$, can be obtained by Laplace inversion of~\eref{eq:Lapx}
\begin{equation}\label{eq:solx}
    x(t) = x_0 \chi(t) + \int_0^{t} dt' \, \varsigma(t-t') \zeta(t'),
\end{equation}
where $\chi(t)$ and $\varsigma(t)$ are the inverse Laplace transforms of the functions $\tilde{\chi}(s)$ and $\tilde{\varsigma}(s)$ defined in \eref{eq:chi} and \eref{eq:sigma}, respectively. Since we are interested in the situation where the particle is in equilibrium with the viscoelastic bath, at all times $t \ge 0$ the stationary probability density function of the particle position $x$ must be the canonical one
\begin{equation}\label{eq:pdfx0}
    P_{eq}(x) = \sqrt{ \frac{\kappa}{2\pi k_B T}} \exp \left( -\frac{\kappa x^2}{2 k_B T} \right).
\end{equation}
\Eref{eq:pdfx0} is consistent with the equipartition relation for the particle position, $x(t)$, at any time $t \ge 0$, namely, $\frac{1}{2}\kappa \left\langle x(t)^2 \right\rangle_{eq} = \frac{1}{2} k_B T $, where $\langle \ldots \rangle_{eq}$ denotes an average with respect to the stationary distribution \eref{eq:pdfx0}. Therefore, from~\eref{eq:solx} and \eref{eq:pdfx0} the following expression can be readily derived for the autocorrelation function of $x(t)$ computed between times $t'$ and $t'+\tau$
\begin{equation}\label{eq:autocorrx}
    \langle x(t'+\tau) x(t') \rangle_{eq} = \langle x(\tau) x(0) \rangle_{eq} =  \frac{k_B T}{\kappa} \chi(\tau),
\end{equation}
where $t', \tau \ge 0$. Note that \eref{eq:autocorrx} shows that the function $\chi(t)$ is proportional to the autocorrelation function of the particle position at thermal equilibrium. Moreover, regardless of the specific form of the memory kernel, $\chi(t)$ generally exhibits the two limit values
\begin{eqnarray}\label{eq:chilim}
        \lim_{t \rightarrow 0} \chi(t) &= \lim_{s\rightarrow \infty} s\tilde{\chi}(s) = 1,\nonumber\\
        \lim_{t \rightarrow \infty} \chi(t) &= \lim_{s\rightarrow 0} s\tilde{\chi}(s) = 0.
\end{eqnarray}
In the second equality of~\eref{eq:chilim}, we have used the fact that $\lim_{s \rightarrow 0} s \tilde{\Gamma}(s) = \lim_{t \rightarrow \infty} \Gamma(t) = 0$ because the stress relaxation modulus of a viscoelastic fluid satisfies $G(t \rightarrow \infty) = 0$. Furthermore, by computing the average of the square of each side of \eref{eq:solx} over an infinite number of realizations of the thermal noise $\zeta(t')$ for a fixed value of $x_0$, then with respect to the stationary distribution $P_{eq}(x_0)$ given in \eref{eq:pdfx0}, and finally multiplying by $\kappa /2$, the use of the equipartition relation leads to the equality 
\begin{equation}\label{eq:squareav}
	\frac{k_B T}{2} = \frac{k_B T}{2} \chi(t)^2 + \frac{\kappa}{2}\int_0^t dt' \int_0^t dt''  \left\langle \zeta(t') \zeta(t'') \right\rangle \varsigma(t-t') \varsigma(t-t''),
\end{equation}
where $\langle \ldots \rangle$ denotes the ensemble average over $\zeta(t')$ during the time interval $0 < t' \le t$.

We now proceed to determine the probability density of finding the particle at position $x$ at time $t>0$ provided that it was located at position $x_0$ at time $t=0$, that we denote as $P(x,t|x_0,0)$. As described in~\sref{sec:pdf}, this is necessary for the calculation of the probability density function of the variation of the harmonic potential energy $U(x)$ at two different times. By definition, this conditional probability is 
\begin{eqnarray}\label{eq:condpdfx}
	P(x,t|x_0,0) & = & \langle \delta\left[ x  - x(t)  \right] \rangle_0,\nonumber\\
			& = & \frac{1}{2\pi} \int_{-\infty}^{\infty} dk \, e^{  ik \left[ x - x_0\chi(t) \right]  } \left \langle \exp \left[ -ik \int_0^t dt' \, \varsigma (t-t') \zeta(t') \right] \right \rangle,
\end{eqnarray}
where $\langle \ldots \rangle_0$ denotes an ensemble average over an infinite number of realizations of the thermal noise $\zeta(t')$ during the time interval $0 < t' \le t$ for a fixed initial condition $x(0) = x_0$. In the second line of \eref{eq:condpdfx}, we have used the Fourier representation of the Dirac delta function, $\delta \left[ x -x(t) \right] = \frac{1}{2\pi} \int_{-\infty}^{\infty} dk\, e^{ik \left[ x - x(t) \right]}$, as well as the expression of the time-domain solution $x(t)$ with initial condition $x(0) = x_0$ given in \eref{eq:solx}. Note that the term $\Phi[-k \varsigma] \equiv \left \langle \exp \left\{ i \int_0^t dt' \, \left[- k \varsigma (t-t') \right] \zeta(t')  \right\} \right \rangle$ in the second line of \eref{eq:condpdfx} can be identified as the characteristic functional of the noise $\zeta(t')$. Since $\zeta(t')$ is assumed to be Gaussian,  $\Phi[-k \varsigma]$ can be simply expressed as~\cite{kubo1991}
\begin{eqnarray}\label{eq:charactfunc}
	\Phi[- k\varsigma] & = & \exp \left[ -\frac{k^2}{2} \int_0^t dt' \int_0^t dt''  \left\langle \zeta(t') \zeta(t'') \right\rangle \varsigma(t-t') \varsigma(t-t'')  \right], \nonumber\\
			& = & \exp \left\{ - k^2 \frac{k_B T}{2\kappa}\left[ 1 - \chi(t)^2 \right] \right\},
\end{eqnarray}
where we have made us of \eref{eq:squareav} in the second line of \eref{eq:charactfunc}. A direct substitution of \eref{eq:charactfunc} into \eref{eq:condpdfx} and a straightforward calculation of the corresponding inverse Fourier transform leads the final expression of the conditional probability density $P(x,t|x_0,0)$
\begin{equation}\label{eq:solSmol}
    P(x,t|x_0,0) = \sqrt{\frac{\kappa}{2\pi k_B T\left[1-\chi(t)^2 \right]}} \exp \left\{ -\frac{\kappa}{2k_BT} \frac{\left[ x - x_0 \chi(t) \right]^2}{1 - \chi(t)^2}\right\}.
\end{equation}
We point out that the same expression for $P(x,t|x_0,0)$ as in \eref{eq:solSmol} can also be obtained by solving a non-Markovian Smoluchowski equation equivalent to the full generalized Langevin equation \eref{eq:GLE} with inertia provided that the system starts from an initial equilibrium state, i.e. with Boltzmann distribution $P_{eq}(x_0, v_0) \propto \exp\left( -\frac{m_{\mathrm{eff}}v_0^2 + \kappa x_0^2}{2 k_B T} \right)$ for the initial conditions $(x_0,v_0)$~\cite{adelman1977,okuyama1986}, and then taking the overdamped limit $m_{\mathrm{eff}} \rightarrow 0$. In either case, the assumption of initial equilibrium at $t = 0$ guarantees the stationarity of the probability density of $x$ at all times $t > 0$, i.e. $\int_{-\infty}^{\infty} dx_0 \, P(x,t|x_0,0) P_{eq}(x_0) = P_{eq}(x)$, as can be verified 
by direct use of \eref{eq:solSmol} with both $P_{eq}(x_0)$ and $P_{eq}(x)$ given by \eref{eq:pdfx0}.}

\section{Probability distribution of the energy exchanged with the bath}\label{sec:pdf}
In this section, we derive an analytical expression for the probability density function of the energy transferred during a given time interval from the viscoelastic fluid, which plays the role of a non-Markovian bath at constant temperature, to the mesoscopic system of interest, which is in this case Brownian particle trapped by the harmonic potential. . 
In situations like this, during a short time interval of duration of at least $dt \sim 10^{-6}$~s, the variation of the \emph{bare} Hamiltonian of the system, $H(x,\dot{x}) = \frac{1}{2}m\dot{x}^2 + U(x) = \frac{1}{2}m\dot{x}^2 + \frac{1}{2} \kappa x^2$, i.e. its energy without taking into account the interaction energy with the bath particles, can be approximated as $dH = (m\ddot{x} + \kappa x) \dot{x} dt \approx \kappa x \dot{x} dt = dU$. This is because for such values of $dt$, the overdamped limit is valid, in such a way that the inertial term $m\ddot{x}$ can be completely neglected with respect to the harmonic force, $-\kappa x$. Furthermore, in absence of external forces performing work on the system, we can interpret the variation $dH = dU$ as the energy randomly exchanged with the viscoelastic fluid during $dt$. If $dH > 0$, the systems absorbs energy from the surroundings, whereas it releases energy if $dH < 0$, thus increasing or decreasing its potential energy, respectively. Therefore, the energy exchanged between the system and the bath during a time interval $[0,\tau]$ at the level of a single stochastic trajectory $x(t)$, with $0 \le t \le \tau$, starting at $x(0) = x_0$ and ending at $ x(\tau) = x_{\tau}$, which we denote as $\mathcal{Q}_{\tau}$, can be determined from
\begin{eqnarray}\label{eq:enert}
    \mathcal{Q}_{\tau} & = & U(x_{\tau}) - U(x_0), \nonumber\\ 
    & = &\frac{\kappa}{2} \left( x_{\tau}^2 - x_0^2\right).
\end{eqnarray}
It should be noted that~\eref{eq:enert} has the same form as the first-law-like relation for the energy balance along a stochastic trajectory of a Brownian particle harmonically trapped in thermal equilibrium in a viscous solvent. In such a  case, the left-hand side represents the \emph{heat} stochastically exchanged with the bath via the fast thermal collisions with the solvent molecules and by viscous friction, and exactly amounts to minus the total energy variation of the environment. However, when the particle is trapped in a viscoelastic fluid, $\mathcal{Q}_{\tau}$ cannot be simply identified as heat since the stochastic energetics of the system also involves the interaction energy between the trapped particle and the surrounding macromolecules suspended in the solvent. The reason of this is that such interactions are not negligible and give rise to temporal correlations that do not relax as quickly as in the case of a purely viscous bath, as manifested by the presence of slowly decaying memory kernels and colored noises
in the coarse-grained description of the particle dynamics provided by the generalized Langevin equation~\eref{eq:GLE}. In this situation, a more detailed description of the energy balance at the system trajectory level could be achieved by introducing the concept of \emph{Hamiltonian of mean force}, which includes the effect of the interaction energy averaged over the degrees of freedom of the bath particles \cite{roux1999,gelin2009,seifert2016}. As discussed in \cite{seifert2016,ding2022}, this approach could eventually allow to properly define the exchanged heat for mesoscopic systems that strongly interact with a bath such as the viscoelastic fluid considered here, even though there is also some criticism against the unambiguous identification of stochastic energy-like quantities other than the applied work under such conditions \cite{talkner2016,talkner2020}. Therefore, in the following, we will not address these ongoing issues on stochastic thermodynamics for systems that are strongly coupled to a heat bath and will only focus on the statistics of $\mathcal{Q}_{\tau}$. In absence of work externally applied to the system, this quantity can be univocally identified as the energy absorbed from or released to the surrounding viscoelastic fluid by the particle, which allows it to explore the harmonic potential beyond the minimum at $x=0$. Indeed, by multiplying the overdamped version of the generalized Langevin equation~\eref{eq:GLE} by a small particle displacement $dx = \dot{x}(t) dt$ performed between times $t \ge 0$ and $t + dt$, and upon integration over $[0,\tau]$ using~\eref{eq:enert}, $\mathcal{Q}_{\tau}$ can be expressed as
\begin{equation}\label{eq:enert1}
    \mathcal{Q}_{\tau} = \int_0^{\tau} dt\, \zeta(t)\dot{x}(t) -\int_0^{\tau} dt \int_0^{t} dt'  \Gamma(t-t') \dot{x}(t') \dot{x}(t), 
\end{equation}
where the first term on the right-hand side corresponds to the energy supplied to the system by the fluctuating thermal force of the bath, $\zeta(t)$, during the stochastic realization of $x(t)$ from $t=0$ to $t = \tau$, whereas the second term represents the energy exchanged by the action of the hydrodynamic drag exerted by the fluid, $-\int_0^{t} dt'  \Gamma(t-t') \dot{x}(t')$, both terms including the effect of the interaction of the trapped particle with the macromolecules of the fluid. \Eref{eq:enert1} correctly reduces to the widely accepted definition of stochastic heat in the case of a purely viscous bath of constant viscosity $\eta$~\cite{sekimoto1998}, in which case $\Gamma(t) = 12 \pi a \eta \delta(t) $, thus yielding
\begin{equation}\label{eq:heatviscous}
    \mathcal{Q}_{\tau} = \int_0^{\tau} dt \, \zeta(t)\dot{x}(t) - \gamma\int_0^{\tau} dt\, \dot{x}(t)^2,
\end{equation}
where $\gamma = 6\pi a \eta$ is the friction coefficient of the particle. Note that the second terms on the right-hand sides of~\eref{eq:heatviscous} and~\eref{eq:enert1} highlight the subtle difference between the properties of the heat exchange that is properly defined for a particle weakly coupled to a viscous bath with respect to the general energy transfer when it is strongly coupled to a viscoelastic one. While in the former case viscous friction always produces energy dissipation into the bath, which corresponds to the negative definite term of~\eref{eq:heatviscous}, in the latter the viscoelastic drag could give rise to either a positive or a negative contribution to the energy transfer because of the long-lived particle interactions comprised in the memory kernel in~\eref{eq:enert1}.

Once the physical interpretation of $\mathcal{Q}_{\tau}$ has been clarified, the calculation of its probability distribution that we denote as $P(\mathcal{Q},\tau)$, can be carried out in a straightforward manner. We realize that, according to~\eref{eq:enert}, $\mathcal{Q}_{\tau}$ is only determined by the difference of the two stochastic variables $x_0^2$ and $x_{\tau}^2$, which are in general correlated for finite values values of $\tau$. Therefore, $P(\mathcal{Q},\tau)$ can be expressed as
\begin{eqnarray}\label{eq:pdfqtau}
P(\mathcal{Q},\tau) & = & \langle \delta \left(  \mathcal{Q}  - \mathcal{Q}_{\tau} \right)\rangle, \nonumber\\
 & = & \int_{-\infty}^{\infty} dx_{\tau} \int_{-\infty}^{\infty} dx_0 P_{eq}(x_0) P(x_{\tau},\tau|x_0,0) \delta \left[ \mathcal{Q} - \frac{\kappa}{2} \left( x_{\tau}^2 - x_0^2\right) \right],
\end{eqnarray}
where $P_{eq}(x_0)$ is the probability density function of the initial position $x_0$, which is given by \eref{eq:pdfx0} since we assume that the system is at all times $t \ge 0$ in thermal equilibrium with the viscoelastic bath, In addition, the function $P(x_{\tau},\tau|x_0,0)$ in~\eref{eq:pdfqtau} corresponds to the probability density of finding the particle at position $x_{\tau}$ at time $\tau > 0$ provided that it was located at $x_0$ at time $t = 0$, whose explicit expression is given in~\eref{eq:solSmol} for any arbitrary memory kernel $\Gamma(t-t')$ governing the stochastic dynamics of $x(t)$. Therefore, by taking into account that the Dirac delta function in the integrand of~\eref{eq:pdfqtau} can be written as
\begin{equation}\label{eq:diracdeltaq}
    \delta \left[ \mathcal{Q} - \frac{\kappa}{2} \left( x_{\tau}^2 - x_0^2\right) \right] = \frac{ \delta\left(x_{\tau} - \sqrt{x_0^2 + \frac{2}{\kappa}\mathcal{Q}} \right) + \delta\left( x_{\tau} + \sqrt{x_0^2 + \frac{2}{\kappa}\mathcal{Q}} \right)}{\kappa \sqrt{x_0^2 + \frac{2}{\kappa}\mathcal{Q}}},
\end{equation}
and by substituting~\eref{eq:diracdeltaq} into~\eref{eq:pdfqtau} in order to integrate over $-\infty < x_{\tau} < \infty$, a more compact expression for $P(\mathcal{Q},\tau)$ can be obtained
\begin{equation}\label{eq:pdfqtau2}
P(\mathcal{Q},\tau) = P_+(\mathcal{Q},\tau) + P_-(\mathcal{Q},\tau),
\end{equation}
where the functions $P_+(\mathcal{Q},\tau)$ and $P_-(\mathcal{Q},\tau)$, defined as
\begin{equation}\label{eq:pdfpm}
    P_{\pm}(\mathcal{Q},\tau) = \frac{1}{\kappa} \int_{-\infty}^{\infty} dx_0 \frac{P_{eq}(x_0) P\left(\pm \sqrt{x_0^2 + \frac{2}{\kappa}\mathcal{Q}},\tau|x_0,0\right)}{\sqrt{x_0^2 + \frac{2}{\kappa}\mathcal{Q}}},
\end{equation}
only involve a single integration over $-\infty < x_0 < \infty$. Moreover, by means of the change of variables 
\begin{equation}\label{eq:changez}
    z_{\pm} = \sqrt{\frac{\kappa}{k_B T}} \left( \sqrt{x_0^2 + \frac{2}{\kappa} \mathcal{Q}} \mp x_0 \right),
\end{equation}
and by introducing the functions \begin{equation}\label{eq:phipm}
    \varphi_{\pm}(\tau) = 1 \pm \chi(\tau), 
\end{equation}
the integrals in~\eref{eq:pdfpm} can be recast as
\begin{equation}\label{eq:pdfpm2}
    P_{\pm}(\mathcal{Q},\tau) = \frac{1}{2\pi k_B T \sqrt{\varphi_+(\tau) \varphi_-(\tau)}} \int_0^{\infty} dz_{\pm} \, \frac{1}{z_{\pm}} \exp\left[-\frac{\left(\frac{\mathcal{Q}}{k_B T}\right)^2}{\varphi_{\pm}(\tau) z_{\pm}^2}  - \frac{z_{\pm}^2}{4 \varphi_{\mp}(\tau) } \right].
\end{equation}
Furthermore, by use of the identity
$\int_0^{\infty} dz \, z^{\nu - 1} e^{ -\beta z^p - \alpha z^{-p} } = \frac{2}{p} \left( \frac{\alpha}{\beta} \right)^{\frac{\nu}{2p}} K_{\frac{\nu}{p}}\left( 2 \sqrt{\alpha \beta} \right)$, where $K_{\frac{\nu}{p}}(2\sqrt{\alpha \beta})$ is the ${\nu}/{p}-$th order modified Bessel function of the second kind with argument $2\sqrt{\alpha \beta}$ \cite{gradshteyn1980}, it can be easily shown that $P_{+}(\mathcal{Q},\tau) = P_{-}(\mathcal{Q},\tau)$, and~\eref{eq:pdfqtau} simplifies to
\begin{equation}\label{eq:pdffinal}
    P(\mathcal{Q},\tau) = \frac{1}{\pi k_B T \sqrt{1-\chi(\tau)^2}} K_0 \left( \frac{1}{\sqrt{1-\chi(\tau)^2}} \frac{|\mathcal{Q}|}{k_B T}\right),
\end{equation}
where $K_0(w) = \int_0^{\infty} du \, \cos \left( w \sinh u  \right)$ denotes the zeroth order modified Bessel function of the second kind. \Eref{eq:pdffinal}, which is the main result of the present paper, provides a surprisingly simple expression valid in the overdamped regime for the probability distribution of the energy exchanged in thermal equilibrium during a time interval of duration $\tau \gtrsim 10^{-6}$~s between the harmonically trapped particle and the surrounding viscoelastic bath.
Note that~\eref{eq:pdffinal} reveals that all the information on the energetic coupling between the particle and the viscoelastic bath is encoded in a single function, namely, $\varphi_+(\tau)\varphi_-(\tau) = 1-\chi(\tau)^2$, whose dependence on $\tau$ is determined by the strength of the harmonic potential and by the specific form of the relaxation modulus of the bath, as will be shown in~\sref{sec:exp} for two specific examples of viscoelastic fluids. Moreover, since the function $\chi(t)$ is proportional to the autocorrelation function of the particle position at thermal equilibrium according to \eref{eq:autocorrx}, the function $\varphi_+(\tau) \varphi_-(\tau)$ contains information of the temporal correlations induced by the energetic coupling of the trapped particle with its non-Markovian environment.

From the explicit expression given by~\eref{eq:pdffinal} and using the definite-integral identity
$\int_0^{\infty} dw \, w^{\lambda-1}K_{\mu}(w) = 2^{\lambda -2} \mathit{\Gamma}(\frac{\lambda-\mu}{2}) \mathit{\Gamma}(\frac{\lambda+\mu}{2})$
for the $\mu-$th order modified Bessel function of the second kind~\footnote{Here, $\mathit{\Gamma}(z) = \int_0^{\infty} du \, u^{z-1} e^{-u}$ represents the well-known gamma function, which must not be confused with the memory kernel $\Gamma(t-t')$ describing the Brownian particle dynamics in~\eref{eq:GLE}.}, all the moments of order $n \ge 0$ of the distribution of $P(\mathcal{Q},\tau)$, i.e.
\begin{equation}\label{eq:qmoment}
    \langle \mathcal{Q}^n_{\tau} \rangle = \int_{-\infty}^{+\infty}d\mathcal{Q} \, P(\mathcal{Q},\tau) \mathcal{Q}^n, 
\end{equation}
can be calculated. Indeed, owing to the symmetry of $P(\mathcal{Q},\tau)$ with respect to $\mathcal{Q} = 0$
\begin{equation}\label{eq:momentsq}
\label{cases}
 \langle \mathcal{Q}^n_{\tau} \rangle =\cases{0,&if $n$ is odd,\\
\left[ \frac{n!}{\left( \frac{n}{2}\right)!}  \right]^2 \left( \frac{k_B T}{2} \right)^n \left[ 1 - \chi(\tau)^2\right]^{\frac{n}{2}},&if $n$ is even.\\}
\end{equation}
In particular, from~\eref{eq:momentsq} we can easily check that $P(\mathcal{Q},\tau)$ is normalized, i.e. $\int_{-\infty}^{\infty} d\mathcal{Q} \, P(\mathcal{Q},\tau) = 1$. Additionally, the mean energy transferred from the bath to the particle is 
\begin{equation}\label{eq:meanq}
    \langle \mathcal{Q}_{\tau} \rangle = 0,
\end{equation}
which simply reflects the fact that detailed balance guarantees the absence of net energy fluxes on average at thermal equilibrium even if the bath is non-Markovian, like the viscoelastic fluid at constant temperature considered here. Moreover, from~\eref{eq:momentsq}, the variance of the exchanged energy during a time interval of duration $\tau$, $ \langle \Delta \mathcal{Q}^2_{\tau}\rangle \equiv \langle \mathcal{Q}^2_{\tau} \rangle - \langle \mathcal{Q}_{\tau}\rangle^2$, is explicitly given by
\begin{equation}\label{eq:varianceq}
   \langle \Delta \mathcal{Q}^2_{\tau}\rangle  = (k_B T)^2 \left[1 - \chi(\tau)^2 \right],
\end{equation}
whose asymptotic value for $\tau \rightarrow \infty$ is $\langle \Delta \mathcal{Q}^2_{\tau \rightarrow \infty}\rangle = (k_B T)^2$, according to the limit values of the function $\chi(t)$ given in~\eref{eq:chilim} for an arbitrary relaxation modulus of the viscoelastic fluid. This result shows that, for sufficiently long intervals over which all temporal correlations induced either by the trapping potential or by the interactions with the medium vanish, the characteristic exchanged energy is $\left( \langle \Delta \mathcal{Q}^2_{\tau \rightarrow \infty} \rangle\right)^{1/2} = \pm k_B T$  regardless of the specific features of the memory kernel and the value of the trap stiffness.
It should be noted that expressions for $\langle \mathcal{Q}_{\tau} \rangle$ and $\langle \Delta \mathcal{Q}_{\tau}^2 \rangle$ that are similar to those given in~\eref{eq:meanq} and~\eref{eq:varianceq} were derived in \cite{chatterjee2009} for the same model investigated here based on the general solution of~\eref{eq:GLE}, $[x(t), \dot{x}(t)]$, for given initial conditions $(x_0,v_0)$ at $t = 0$. However, the calculation of higher-order moments following this approach becomes too involved, thus being feasible in practice only for the determination of the mean and the variance. Here, we overcome this limitation by directly deriving an expression for the probability distribution of $\mathcal{Q}_{\tau}$, thereby having full knowledge of its statistical properties. For instance, from~\eref{eq:momentsq}, the skewness of $P(\mathcal{Q},\tau)$ is $0$, whereas the kurtosis is independent of $\tau$ and has the constant value
\begin{equation}\label{eq:kurt}
    K_{\tau} \equiv \frac{\langle \mathcal{Q}^4_{\tau} \rangle}{ \langle \mathcal{Q}^2_{\tau} \rangle^2}  = 9.
\end{equation}

Some additional remarks can be made about~\eref{eq:pdffinal}. For instance, it can be easily checked that it correctly reduces to the well know expression for the probability distribution of the heat exchanged between a spherical particle of radius $a$ confined by a harmonic potential and a viscous heat bath of constant viscosity $\eta$. Indeed, by taking the instantaneous memory kernel $\Gamma(t - t') = 2 \gamma \delta(t -  t')$, which corresponds to a time-independent friction coefficient $\gamma = 6 \pi a \eta$, the functions in~\eref{eq:ODGamma} and~\eref{eq:chi} become $\tilde{\Gamma}(s) = \gamma$, and $\tilde{\chi}(s) = \left(s + \frac{1}{\tau_{\gamma}} \right)^{-1}$, respectively, where $\tau_{\gamma} = \gamma/\kappa$ is the characteristic relaxation time of the particle that originates from the balance between viscous friction and the restoring force exerted by the harmonic potential. In this case, the inverse Laplace transform of $\tilde{\chi}(s)$ is $\chi(t) = \exp\left(-\frac{ t}{\tau_{\gamma}}\right)$, so that $\varphi_+(\tau) \varphi_-(\tau) = 1 - e^{-2\tau/\tau_{\gamma}}=2 e^{-\tau/\tau_{\gamma}} \sinh (\tau / \tau_{\gamma})$, thus leading to the following expression for the probability density function of the exchanged heat
\begin{equation}\label{eq:pdfviscous}
    P(\mathcal{Q},\tau) =  \frac{e^{\frac{\tau}{2 \tau_{\gamma}}}}{\pi k_B T \sqrt{2 \sinh \left( \frac{\tau}{\tau_{\gamma}}\right)}} K_0 \left(  \frac{|\mathcal{Q}|}{k_B T} \sqrt{\frac{1}{2} \left[ 1+\coth \left( \frac{\tau}{\tau_{\gamma}} \right) \right] } \right),
\end{equation}
which is the same formula as previously derived in~\cite{chatterjee2010} using path integrals for the Markovian dynamics corresponding to the Langevin equation~\eref{eq:Lang}.
Another interesting property of~\eref{eq:pdffinal} is that, irrespective of the specific details of the fluid viscoelastic properties and of the strength of the trapping potential, in the long-time limit $\tau \rightarrow \infty$ it tends to 
\begin{equation}\label{eq:pdfqlong}
    P(\mathcal{Q},\tau\rightarrow \infty) = \frac{1}{\pi k_B T } K_0 \left( \frac{|\mathcal{Q}|}{k_B T} \right),
\end{equation}
where we have used the asymptotic value $\chi(t \rightarrow \infty) = 0$ of the function $\chi(t)$ given in~\eref{eq:chilim}. It is noteworthy that~\eref{eq:pdfqlong} is identical to the formula derived in \cite{imparato2007} for long-time behavior of the probability distribution of the heat exchanged between a Brownian particle trapped by a stationary harmonic potential and a viscous fluid bath. In addition, asymptotic expressions similar to~\eref{eq:pdfqlong} have also been derived for the heat probability distribution of harmonically trapped particles in non-stationary~\cite{gomezsolano2011,crisanti2017} and active media~\cite{goswami2019}, for particles confined by non-linear potentials at low temperature~\cite{fogedby2009}, as well as for the fluctuations of the energy current through harmonic chains connected to two active reservoirs \cite{sarkar2023}. This observation suggests that the zeroth order modified Bessel function of the second kind captures the asymptotic behavior of the energy exchange process with the environment of a broad variety of overdamped Brownian systems  provided that the bath exhibits fluid-like behavior in the long-time limit.

\section{Experimental results}\label{sec:exp}

In this section, we experimentally verify the main results that were previously derived in~\sref{sec:pdf}. To this end, we use as viscoelastic baths two kinds of complex fluids with distinct rheological properties. The first fluid is a worm-like micellar solution, which is composed of the surfactant cetyltrimethylammonium bromide (CTAB) and the salt sodium salycilate (NaSal) at equimolar concentration of 2.5 mM in ultrapure water (resistivity 18.2~M$\Omega$~cm at $25^{\circ}$C). At this concentration, which is above the critical micelle concentration of pure CTAB (0.83~mM), the surfactant molecules self-organize in
flexible cylindrical micelles that have a radius of $\sim 2$~nm, an average contour length of $\sim 200$~nm, and a persistence length of $\sim 50$~nm \cite{lam2019}. The second fluid
consist of a solution of the water-soluble polymer polyethylene oxide (PEO, molecular weight $M_w=4\times 10^6$~Da) dissolved in ultrapure water at a concentration of $c = 0.1$~\%wt, which is smaller than the critical overlap concentration ($c^* \approx 0.65$~\%wt) and corresponds to the semi-diluted particle regime in which polymer chains are non-entangled but inter-chain interactions are not negligible~\cite{ebagninin2009}. For the sake of simplicity, in the following these fluids will be simply referred to as \emph{micellar solution} and \emph{polymer solution}, respectively.  A complete homogenization of both solutions
is achieved by continuous stirring over 24~h, thus becoming transparent to visible light and endowed with viscoelasticity due to the presence of the two types of macromolecules that were previously described.

\begin{figure}
\includegraphics[width=0.85\columnwidth]{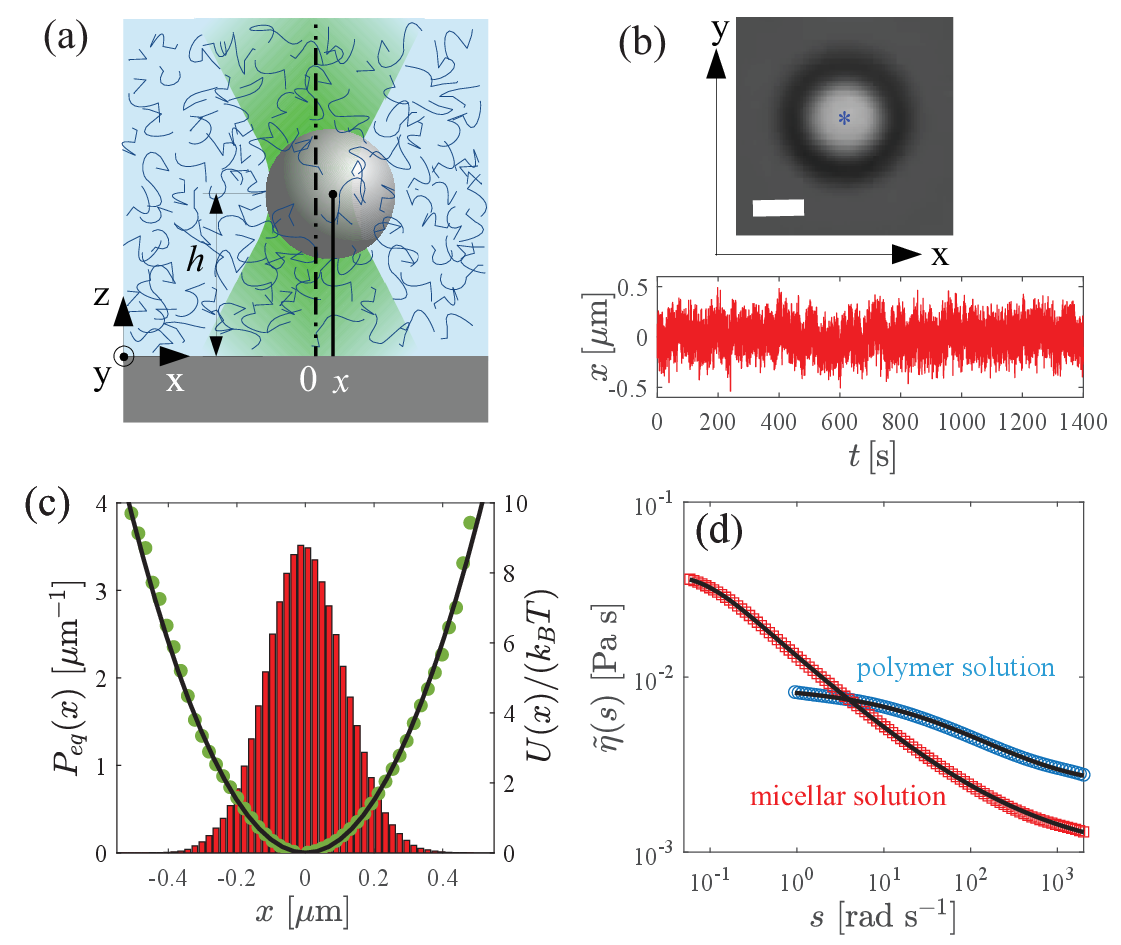}
\caption{(a) Schematic representation of a spherical bead of radius $a$, trapped at a distance $h$ above the lower wall of the sample cell containing a viscoelastic fluid, which is achieved by means of optical tweezers realized by a tightly focused laser beam (green area). The corresponding coordinate system used in the data analysis is also sketched. (b) Upper panel: snapshot in the x-y plane of a bead trapped in the micellar solution, with the location of its center of mass $(x,y)$ marked as an asterisk. The horizontal scale bar represents 1~$\mu$m. Lower panel: example of the stochastic time evolution of the coordinate $x$ of the same particle trapped in the micellar solution. (c) Probability density function of the particle position, $x$, in thermal equilibrium with the micellar solution (vertical bars), experimental profile of the optical trapping potential ($\circ$),
and corresponding quadratic fit (solid line), from which the trap stiffness, $\kappa$, is calculated. (d) Frequency-dependent viscosities of the polymer ($\circ$) and the micellar ($\square$) solutions, which are determined experimentally from the particle motion. Solid lines represent the best non-linear fitting of the experimental data to the model given in~\eref{eq:CYmodel}.}\label{fig:1}
\end{figure}

Once the homogeneous micellar and polymer solutions are prepared, a small amount of spherical silica beads (radius $a = 1\,\mu$m) are dispersed in them, thus resulting in viscoelastic fluid suspensions at very low concentration (less than $1$ colloidal bead in 1 nl of solution). Sample cells made of a microscope glass slide and a coverslip stuck together parallel to each other by double-sided adhesive tape (thickness $\sim 100\,\mu$m), are filled with the viscoelastic suspension of interest, and sealed with epoxy glue in order to prevent leakage and evaporation of the fluid. Then, a green Gaussian laser beam (wavelenght 532~nm) is tightly focused inside the sample cell using an oil-immersion objective ($100 \times$, numerical aperture NA~=~1.3), which allows one to trap a single bead at $h \approx 20\,\mu$m above the lower solid wall of the cell, as sketched in~\fref{fig:1}(a). Such an optical trap confines the stochastic particle motion within a harmonic potential, whose stiffness $\kappa$ is kept constant for each fluid by fixing the laser power. Videos of the trapped particle are recorded using a CMOS camera at a sampling rate of $f_s = 2000$ frames per second during approximately $t_{max} = 25$~min. The position $(x, y)$ of the center of mass of two-dimensional projection in the x-y plane of the particle is detected  with a spatial resolution of 5~nm using standard particle-tracking routines, as illustrated by the snapshot shown in the upper panel of~\fref{fig:1}(b). The measurements are performed at temperatures of $T = 23.5 \pm 0.2^{\circ}$C and $T = 21.0 \pm 0.2^{\circ}$C for the micellar and polymer solutions, respectively.

In the lower panel of~\fref{fig:1}(b) we show an example of the stochastic time evolution of a single coordinate, e.g. $x(t)$, of a particle trapped in the micellar solution, which is  recorded over approximately 25 minutes. The trap stiffness can be computed by, e.g. a quadratic-polynomial fit of the potential $U(x) = \frac{1}{2} \kappa x^2$ that is obtained from the Boltzmann distribution of the particle position given by~\eref{eq:pdfx0}, 
as shown in~\fref{fig:1}(c). An alternative method based on the equipartition relation for $x$ is described in~\ref{sec:microrheology}, and leads to consistent values for $\kappa$. The same procedure is followed to characterize the trapping potential in the case of the polymer solution. Also, as described in~\ref{sec:microrheology}, by means of the numerical calculation of the Laplace transform of the mean-squared displacement of $x(t)$, we can determine the Laplace-frequency dependent viscosity of the fluid, $\tilde{\eta}(s)$, where we only consider positive real frequencies, $s > 0$,  for the sake of clarity. In~\fref{fig:1}(d) we plot the dependence of $\tilde{\eta}(s)$ on $s$ for both the the micellar and the polymer solution. We note that in both cases, such dependencies can be fitted to the function
\begin{equation}\label{eq:CYmodel}
    \tilde{\eta}(s) = \eta_{\infty} + \frac{\eta_0 - \eta_{\infty}}{\left[1 + (\lambda s)^{\alpha}\right]^{\frac{1-n}{\alpha}}},
\end{equation}
whose functional form is consistent with the so-called Carreau-Yasuda equation that is commonly used as a phenomenological model for the shear-rate dependent viscosity of non-Newtonian fluids~\cite{yasuda1981}. In~\eref{eq:CYmodel}, $\eta_0$ and $\eta_{\infty}$ correspond to the values of the fluid viscosity at vanishing and infinite frequencies, respectively, $\lambda$ is the characteristic relaxation time of the fluid, whereas $\alpha \ge 0$ and $n \le 1$ are exponents that depend on the specific microscopic features of the fluid and describe a power-law-like decay of $\tilde{\eta}(s) - \eta_{\infty} \propto s^{-(1-n)}$ at intermediate frequencies. The values of the parameters $\eta_0$, $\eta_{\infty}$, $\lambda$, $\alpha$ and $n$ for each fluid along with the  values of $\kappa$ for the trapping harmonic potentials in each experiment are listed in~\tref{tab:param}. Note that in the case of the polymer solution, we have restricted the outer exponent in~\eref{eq:CYmodel} to the value $(1-n)/\alpha = 1$ by setting $n = 1-\alpha$, which corresponds to the so-called Cross rheological model characterized by a single fitting exponent $\alpha$~\cite{cross1979}. This choice is motivated by previous rheological studies on the macroscopic non-Newtonian behavior of semidilute PEO polymeric solutions reported in the literature~\cite{ebagninin2009}.
On the other hand, both $n$ and $\alpha$ have been employed as independent free fitting parameters for the micellar solution. It should also be noted in~\fref{fig:1}(d) that the dependence of the viscosity on frequency is much more pronounced for the micellar solution than for the polymer solution, where the characteristic relaxation time $\lambda$ of the former is three orders of magnitude larger than that of the latter. Such a marked difference in the rheological properties of these fluids allows us to analyze the stochastic energy exchange of the trapped particle with two types of non-Markovian baths, one with rather strong viscoelastic behavior (the micellar solution) and the other with much weaker viscoelasticity (the polymer solution).

\begin{table}
\caption{\label{tab:param}Parameters characterizing the fluid viscoelasticity and the optical harmonic potential for each experiment.}
\begin{indented}
\item[]\begin{tabular}{@{}lcccccc}
\br
Viscoelastic fluid & $\eta_0$~[Pa~s] & $\eta_{\infty}$~[Pa~s]& $\lambda$~[s] & $\alpha$ & $\frac{1-n}{\alpha}$ & $\kappa$~[N~m$^{-1}$]\\
\mr
\emph{Micellar solution}& 0.0391 & 0.0009 & 11.910 & 2.125 & 0.216 & $ 3.08\times 10^{-7}$\\
\emph{Polymer solution}&0.0090&0.0017&0.022 &0.533 & 1 & $1.43\times 10^{-7}$\\
\br
\end{tabular}
\end{indented}
\end{table}

\begin{figure}
\includegraphics[width=0.9\columnwidth]{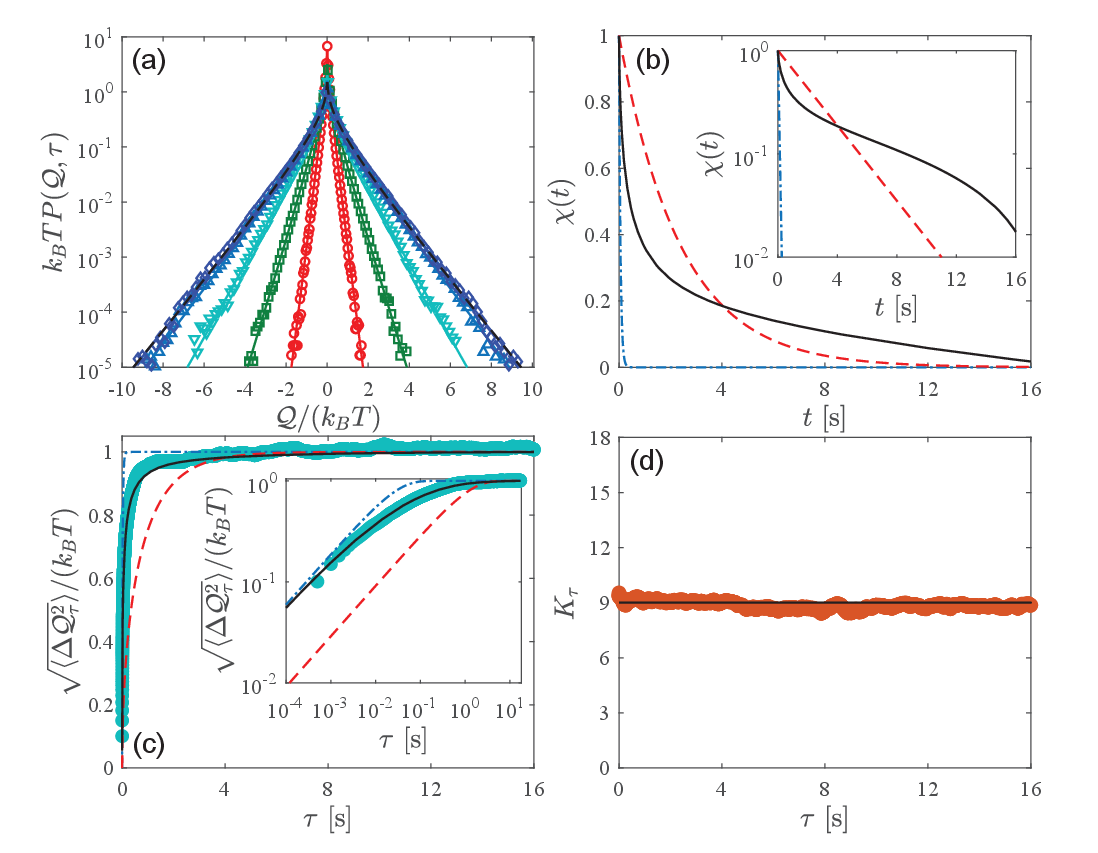}
\caption{(a) Probability density function of the energy exchanged between a colloidal bead trapped by a harmonic optical potential ($\kappa = 3.08 \times 10^{-7}$~N~m$^{-1}$) and the surrounding micellar solution, over time intervals of different duration: $\tau = 0.001$~s ($\circ$), $\tau = 0.01$~s ($\square$), $\tau = 0.1$~s ($\triangledown$), $\tau = 1$~s ($\triangle$), and $\tau = 10$~s ($\diamond$). Solid lines represent the analytical expression given in~\eref{eq:pdffinal} for the corresponding values of $\tau$, whereas the black dashed line depicts the limiting expression for $\tau \rightarrow \infty $ given by~\eref{eq:pdfqlong}. (b) Dependence of $\chi(t)$ on time $t$ for the same system (solid line), which was obtained by numerical inversion of~\eref{eq:chi} using the functional form~\eref{eq:CYmodel} for the frequency-dependent viscosity of the micellar solution shown in~\fref{fig:1}(d). The dashed and dotted-dashed lines represent $\chi(t)$ for a particle in Newtonian fluids of viscosities $\eta_0 = 0.0391$~Pa~s and $\eta_{\infty} = 0.0009$~Pa~s, respectively. Inset: semi-log representation of the main plot. (c) Variance of the experimentally-determined energy exchange as a function of the duration of the time interval over which it takes place ($\circ$). The solid line represents the theoretical prediction of~\eref{eq:varianceq} using the numerical curve plotted as a solid line in~\fref{fig:2}(b), whereas
the dashed and dotted-dashed lines correspond to the variances for a particle in Newtonian fluids of viscosities  $\eta_0 = 0.0391$~Pa~s and $\eta_{\infty} = 0.0009$~Pa~s, respectively. Inset: double-log representation of the main plot.
(d) Kurtosis of the energy exchange distribution as a function of the duration of the time interval ($\circ$). The solid line depicts the constant value $K_{\tau} = 9$ predicted by~\eref{eq:kurt}.}\label{fig:2}
\end{figure}

From the experimental particle trajectories, we now proceed to compute the probability density function of the energy exchanged between the trapped particle and its corresponding viscoelastic surroundings, as well as its respective standard deviation and kurtosis for time intervals of different duration $\tau$, typically  $f_s^{-1} \le \tau \lesssim 10$~s, where $f_s^{-1} = 5\times 10^{-4}$~s. For this purpose, for a fixed value of $\tau$ we directly use~\eref{eq:enert} to compute stochastic values of $\mathcal{Q}_{\tau}$, whose normalized histogram allow us to determine the probability density function $P(\mathcal{Q},\tau)$. Since the system is at thermal equilibrium, its stationarity allows us to choose the initial time in~\eref{eq:enert} as any time $t'$ along the recorded trajectory such that $0 \le t' \le t_{max} - \tau$, i.e. $\mathcal{Q}_{\tau} = \frac{1}{2}\kappa \left( x_{t'+\tau}^2 - x_{t'}^2 \right)$. In this way, the probability density function $P(\mathcal{Q},\tau)$ and its corresponding moments are computed over approximately $3\times10^6$ data points of the random variable $\mathcal{Q}_{\tau}$ for each experiment.

In~\fref{fig:2}(a) we show the results for the micro-bead trapped in the micellar solution, where we verify that the probability density function of $\mathcal{Q}_{\tau}$ is accurately described by the zeroth order modified Bessel function of the second kind given in~\eref{eq:pdffinal} over six orders of magnitude of $P(\mathcal{Q},\tau)$ for all the explored values of $\tau$ spanning 5 order of magnitude ($10^{-3} -$10~s). To this end, for each value of the duration $\tau$ the value of $\chi(\tau)$ is computed by a numerical Laplace inversion of the expression given in~\eref{eq:chi} by means of the Talbot method~\cite{talbot1979}, where we use the experimental value of the trap stiffness as well as the functional form for $\tilde{\eta}(s)$ given by~\eref{eq:CYmodel}, with the corresponding values of its parameters: $\eta_0 = 0.0391$~Pa~s, $\eta_{\infty} = 0.0009$~Pa~s, $\lambda = 11.910$~s, $\alpha = 2.125$, and $n = 0.541$. We also verify that, as $\tau$ increases, the experimental shape of $P(\mathcal{Q},\tau)$ becomes independent of $\tau$ and tend to the function given in~\eref{eq:pdfqlong}, as can be checked in~\fref{fig:2}(a) for $\tau \ge 10$~s, at which the curve of $P(\mathcal{Q},\tau)$ is practically indistinguishable from  $P(\mathcal{Q},\tau \rightarrow \infty)$. In~\fref{fig:2}(b) we plot the dependence of the function $\chi(t)$ on time $t$, which was numerically determined as previously described using the Talbot method. For comparison, we also plot the behavior of $\chi(t)$ in the case of Newtonian fluids with viscosities $\eta_{\infty}$ and $\eta_0$ for the same value of the trap stiffness. We observe that, unlike in a Newtonian fluid, where $\chi(t)$ is merely an exponential decay with characteristic time given by the ratio between the corresponding friction coefficient of the particle and the trap stiffness, in the micellar solution $\chi(t)$ exhibits an intricate non-mono-exponential behavior as a function of $t$. For instance, for the specific value of $\kappa$ used in the experiment, in the micellar solution $\chi(t)$ decays slower than in the Newtonian fluid with viscosity $\eta_{\infty}$ but faster than in the case $\eta_{0}$ up to $t \approx 4$~s. However, for $t \gtrsim 4$~s, the non-Markovian nature of the micellar fluid bath gives rise to a very slow decay of $\chi(t)$ that becomes even slower than the exponential decay $e^{-\kappa t/(6\pi a \eta_0 )}$, then finally approaching $\chi(t) \rightarrow 0$ as expected in any fluid as $t \rightarrow \infty$. To highlight the role of the viscoelasticity of the micellar solution on the energy exchange process with the particle, in~\fref{fig:2}(c) we plot as symbols the experimentally-determined standard deviation of $\mathcal{Q}_{\tau}$, $\sqrt{\langle \Delta \mathcal{Q}_{\tau}^2\rangle}$, as a function of the time interval duration $\tau$. We verify that the square root of the theoretical expression of the variance given in~\eref{eq:varianceq}, which is computed using the numerical curve of $\chi(t)$ plotted in~\fref{fig:2}(b), agrees very well with the experimental data, as shown by the solid line in~\fref{fig:2}(c) for all the values of $\tau$ investigated here. In the same plot, we can also check experimentally that the standard deviation of the energy exchange converges to the limit value $k_B T$ as $\tau \rightarrow \infty$, in accord with the expected asymptotic properties of the energy exchange fluctuations derived at thermal equilibrium from the generalized Langevin model~\eref{eq:GLE}.
Once again, for comparison purposes, we also plot in~\fref{fig:2}(c) the theoretical predictions of~\eref{eq:varianceq} for the standard deviation of $\mathcal{Q}_{\tau}$ as a function of $\tau$ for the heat exchange in Newtonian fluids of viscosities $\eta_{0}$ (dashed line) and $\eta_{\infty}$ (dotted-dashed line). We find that, for $0 < \tau \lesssim 4$~s, the behavior of $\sqrt{\langle \Delta \mathcal{Q}_{\tau}^2\rangle}$ in the micellar solution is comprised between those in Newtonian fluids of viscosities $\eta_{0}$ and $\eta_{\infty}$, while for $\tau \gtrsim 4$~s, the convergence to the limit value $k_B T$ becomes slower in the viscoelastic micellar fluid than in both Newtonian cases. Interestingly, we observe in the inset of~\fref{fig:2}(c) that even though none of these Newtonian curves describes correctly the full dependence of the standard deviation of $\mathcal{Q}_{\tau}$ on $\tau$ in the micellar solution, at small values of $\tau$ its behavior is very close that in a Newtonian fluid with viscosity $\eta_{\infty}$, i.e. $\sqrt{\langle \Delta \mathcal{Q}_{\tau}^2 \rangle} \approx k_B T \left[ 1 - e^{-2 \kappa \tau / (6\pi a \eta_{\infty})}\right]$. This observation can be explained by the fact that, during sufficiently short time intervals, the energy exchange process between the particle and the viscoelastic bath is dominated by the random heat transfer due to the fast collisions of the solvent molecules, which is characterized by the high-frequency friction coefficient $6\pi a \eta_{\infty}$. For longer time intervals, interactions between the trapped bead and the surrounding wormlike micelles become important, thus leading to a transient energy storage in the micelles, which translates into systematic deviations of the standard deviation of $\mathcal{Q}_{\tau}$ from the Newtonian curve with viscosity $\eta_{\infty}$. As $\tau$ becomes much larger than the characteristic correlation times encoded in the behavior of $\chi(t)$, the stochastic energy exchange becomes independent of both the trapping potential and the non-Markovian properties of the viscoelastic micellar bath, thereby resulting in the convergence of the standard deviation of $\mathcal{Q}_{\tau}$ to the characteristic scale of thermal energy, $k_B T$. We point out that, even though we verify experimentally that the standard deviation of $\mathcal{Q}_{\tau}$ depends on $\tau$ through the function $\chi(\tau)$, the kustosis remains independent of $\tau$, very close to the constant value $K_{\tau} = 9$ that is predicted by~\eref{eq:kurt}, as shown in~\fref{fig:2}(d). This confirms that for all the values of $\tau$ analyzed here, the experimental shape of $P(\mathcal{Q},\tau)$ is precisely described by the zeroth order modified Bessel function of the second kind given in~\eref{eq:pdffinal}.

\begin{figure}
\includegraphics[width=0.95\columnwidth]{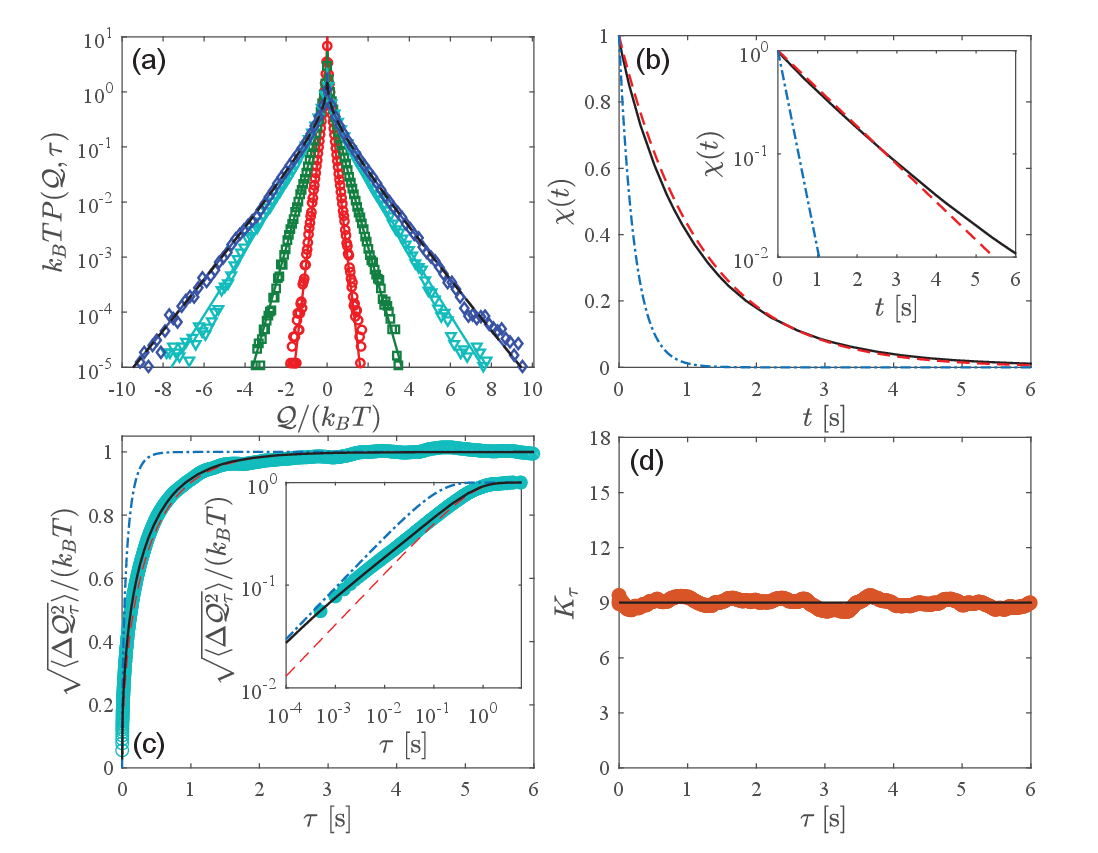}
\caption{(a) Probability density function of the energy exchanged between a colloidal bead trapped by a harmonic optical potential ($\kappa = 1.43 \times 10^{-7}$~N~m$^{-1}$) and the surrounding polymer solution, over time intervals of different duration: $\tau = 0.005$~s ($\circ$), $\tau = 0.05$~s ($\square$), $\tau = 0.5$~s ($\triangledown$), and $\tau = 5$~s ($\triangle$). Solid lines represent the analytical expression given in~\eref{eq:pdffinal} for the corresponding values of $\tau$, whereas the dashed line depicts the limiting expression for $\tau \rightarrow \infty $ given by~\eref{eq:pdfqlong}. (b) Dependence of $\chi(t)$ on time $t$ for the same system (solid line), which was obtained by numerical inversion of~\eref{eq:chi} using the functional form \eref{eq:CYmodel} for the frequency-dependent viscosity of the polymer solution shown in~\fref{fig:1}(d). The dashed and dotted-dashed lines represent $\chi(t)$ for a particle in Newtonian fluids of viscosities $\eta_0 = 0.0090$~Pa~s and $\eta_{\infty} = 0.0017$~Pa~s, respectively. Inset: semi-log representation of the main plot. (c) Variance of the experimentally-determined energy exchange as a function of the duration of the time interval over which it takes place ($\circ$). The solid line represents the theoretical prediction of~\eref{eq:varianceq} using the numerical curve plotted as a solid line in~\fref{fig:3}(b), whereas
the dashed and dotted-dashed lines correspond to the variances for a particle in Newtonian fluids of viscosities $\eta_0 = 0.0090$~Pa~s and $\eta_{\infty} = 0.0017$~Pa~s, respectively. Inset: double-log representation of the main plot.
(d) Kurtosis of the energy exchange distribution as a function of the duration of the time interval ($\circ$). The solid line depicts the constant value $K_{\tau} = 9$ predicted by~\eref{eq:kurt}.}\label{fig:3}
\end{figure}

The results for the probability density function of the energy exchange in the polymer solution are shown in~\fref{fig:3}(a). In the case of this weakly viscoelastic fluid, we also corroborate that the theoretical expression for this probability distribution given in~\eref{eq:pdffinal} is in excellent agreement with the experimental results over six orders in magnitude for the values of $P(\mathcal{Q},{\tau})$ and at least 4 orders of magnitude in $\tau$. We find that for $\tau \ge 5$~s, $P(\mathcal{Q},\tau)$ has already converged to the $\tau$-independent function given in~\eref{eq:pdfqlong}, see the {\color{red}black} dashed line in~\fref{fig:3}(a), in which case all the temporal correlations in the energy exchange process induced by the trapping potential and the bath vanish. In~\fref{fig:3}(b) we plot the behavior of the function $\chi(t)$ for the particle trapped in the polymer solution (solid line), and the hypothetical cases of the same particle trapped in Newtonian  fluids with viscosities $\eta_0 = 0.0090$~Pa~s (dashed line), and $\eta_{\infty} = 0.0017$~Pa~s (dotted-dashed line). We observe that, because of the weak viscoelastic behavior of the polymer solution manifested through the very smooth frequency-dependence of its viscosity $\tilde{\eta}(s)$ plotted in~\fref{fig:1}(d), the dependence of $\chi(t)$ on $t$ in the viscoelastic case is rather close to $e^{-\kappa/(6\pi a \eta_0)}$. However, the non-Markovianity of the bath induced by presence of the suspended polymers gives rise to deviations from such a mono-exponential decay, which can be seen upon closer inspection of this plot and of the semi-log representation shown in the inset of~\fref{fig:3}(b). Indeed, it turns out that the behavior of $\chi(t)$ in the polymer solution is qualitatively similar to that in the micellar fluid. Specifically, at short times $t \lesssim 2.6$~s, the values of $\chi(t)$ in the polymer solution lie between those in Newtonian fluids of viscosities $\eta_{\infty}$ and $\eta_0$, whereas they exceed both Newtonian cases for $t \gtrsim 2.6$~s, thus revealing the presence of slowly decaying correlations in the dynamics of the trapped particle even if the fluid bath is only weakly viscoelastic. Then, using the numerical curve $\chi(t)$ shown in~\fref{fig:3}(b), we can calculate the dependence of the standard deviation of $\mathcal{Q}_{\tau}$, $\sqrt{\langle \Delta \mathcal{Q}_{\tau}^2 \rangle}$, on the time interval duration $\tau$ given by~\eref{eq:varianceq}, which is plotted as a solid line in~\fref{fig:3}(c). Again, we observe a conspicuous agreement between this prediction and the standard deviation of the energy exchange computed from the experimental data points for values of $\tau$ spanning five orders of magnitude ($5\times 10^{-4} - 6$~s). Although the viscoelasticity of the polymer solution is not as pronounced as that of the micellar one, we are able to detect similar qualitative features of the dependence of $\sqrt{\langle \Delta \mathcal{Q}_{\tau}^2 \rangle}$ on $\tau$. In particular, we find in the inset of~\fref{fig:3}(c) that at sufficiently small values of $\tau$,  $\sqrt{\langle \Delta \mathcal{Q}_{\tau}^2 \rangle} \approx k_B T \left( 1 - e^{-2 \kappa \tau / (6\pi a \eta_{\infty})}\right)$
due to the dominance of the heat exchange with the solvent molecules, followed by an intermediate regime where deviations from the behavior in a Newtonian fluid occur due to the slow temporal correlations induced by interactions with the polymer chains, whereas for sufficiently large $\tau$ we observe the convergence to the asymptotic value $\sqrt{\langle \Delta \mathcal{Q}_{\tau \rightarrow \infty}^2 \rangle} = k_B T$. Finally, in~\fref{fig:3}(c) we verify that the values of the kurtosis of the experimentally determined distributions for the energy exchange in the polymer solution are very close to the theoretical one $K_{\tau} = 9$.

\section{Summary and concluding remarks}
\label{sect:conclude}

In this paper, we have investigated theoretically and experimentally the statistics of the stochastic energetics of an overdamped Brownian particle trapped at constant temperature in a viscoelastic fluid acting as a non-Markovian thermal reservoir. By analyzing the simplest nontrivial situation where the trapping potential is harmonic and no work is applied to the system but the fluid has an arbitrary relaxation modulus, we were able to derive an analytical expression for the probability density function of the energy exchanged between the particle and its surroundings during a finite time interval. We have shown that this probability distribution, which can be written as a modified Bessel function of the second kind with even symmetry, as well as all its even moments, can be expressed in terms of a function that depends on the time-interval duration and is proportional to the squared autocorrelation function of the particle position at thermal equilibrium. Such expressions include the effect of temporal correlations induced by both the confining potential as well as the interaction with the bath particles, thereby representing an extension of previously derived formulae for the heat exchange of Brownian systems in Markovian baths, e.g. viscous fluids. Our results are clearly illustrated in experiments using colloidal particles optically trapped in two types of viscoelastic fluids with distinc rheological and structural properties, namely, a wormlike micellar solution and a semidilute polymer solution, which are of widespread interest in soft matter systems.

In recent years, there has been a growing interest in extending the theoretical framework of stochastic thermodynamics beyond the usual assumption of weak coupling with a heat bath for small systems both in the classical and quantum realms, e.g. mesoscopic systems that interact with a non-Markovian bath like the viscoelastic one investigated here. Although the thermodynamic description of such systems subject to external driving forces are a focus of attention in various theoretical investigations~~\cite{seifert2016,ding2022,talkner2020,strasberg2017,aurell2018,cockrell2022,venturelli2022}, very few experimental studies have been conducted in a controlled fashion even close to thermal equilibrium~\cite{carberry2007,huang2021,narinder2021}. Therefore, the results presented in this paper represent a first step towards a quantitative characterization of stochastic energy exchanges of Brownian systems in non-Markovian environments and their connection with the slowly-decaying correlations that emerge due to the interactions with the bath particles. Further experimental and theoretical efforts based on this model system could help to investigate other relevant aspects on the stochastic energetics at strong coupling under more intricate conditions, e.g. in thermally-activated escape processes over energetic barriers with memory friction~\cite{ferrer2021}, the efficiency of cyclic Brownian engines operating in non-Newtonian fluids~\cite{krishnamurthy2016,guevara2023,nalupurackal2023}, and work fluctuations of colloids periodically or stochastically driven in viscoelastic media~\cite{kundu2021,das2023}.

\section*{Acknowledgments}
We acknowledge support from DGAPA-UNAM PAPIIT Grant No. IA104922. 

\appendix

\section{Determination of the frequency-dependent viscosity}\label{sec:microrheology}

In this Appendix, we provide more details on the experimental determination of the Laplace-frequency dependent viscosity, $\tilde{\eta}(s)$, of the two distinct viscoelastic fluids used as non-Markovian baths, which in turn determines the behavior of the memory kernel $\Gamma(t-t')$ experienced by the embedded colloidal bead in the overdamped limit according to~\eref{eq:ODGamma}. 
From the expression given in \eref{eq:solx} for the solution $x(t)$ of the particle position at time $t > 0$, the corresponding mean-squared displacement can be computed between times $t'$ and $t'+\tau$
\begin{equation}\label{eq:msdx}
    \langle \Delta x(\tau)^2 \rangle \equiv \langle \left[ x(t'+\tau) - x(t') \right]^2 \rangle = \frac{2k_B T}{\kappa} \left[ 1-\chi(\tau) \right],
\end{equation}
where $t', \tau \ge 0$.
Note that $\langle \Delta x(\tau)^2 \rangle$, can be directly computed from the experimentally measured trajectories of the trapped particle, and since it is implicitly related to the memory kernel through the function $\chi(\tau)$ on the right-hand sides of~\eref{eq:msdx}, the latter can be exploited to extract the viscoelastic properties of the investigated fluids. For instance, from~\eref{eq:ODGamma} and~\eref{eq:msdx}, it can be readily shown that the Laplace-frequency-dependent viscosity of the fluid can be expressed in the overdamped limit in terms of the Laplace transform of the mean-squared displacement of the particle position, $\widetilde{\langle \Delta x(s)^2\rangle} = \int_0^{\infty} d\tau\, e^{-s \tau} \langle \Delta x(\tau)^2\rangle$, by means of the generalized Stokes-Einstein relation
\begin{equation}\label{eq:GSER}
    \tilde{\eta}(s) = \frac{1}{6\pi a s} \left[ \frac{2k_B T}{s \widetilde{\langle \Delta x(s)^2\rangle}} - \kappa \right].
\end{equation} 
\Eref{eq:GSER} allows one to find the general dependence of $\tilde{\eta}(s)$ on $s \in \mathbb{C}$, in particular when $s$ is purely real,  provided that the value of the harmonic trap stiffness, $\kappa$, and the specific form of $\widetilde{\langle \Delta x(s)^2\rangle}$, are both known. In practice, the value of $\kappa$ can be simply determined from the autocorrelation function given in~\eref{eq:autocorrx} by taking $\tau = 0$, i.e.
\begin{equation}\label{eq:kap}\kappa = \frac{k_B T}{\langle x(t)^2 \rangle}_t. 
\end{equation}
In~\eref{eq:kap}, $\langle \ldots \rangle_t$ denotes a time average over the temporal evolution of a single stochastic particle trajectory of duration $t_{max}$ much larger than the characteristic relaxation times of the system, $\{ x(t), 0 \le t \le t_{max} \}$, which can be computed instead of an ensemble average over independent initial conditions $x_0$ due to the stationarity of the system at thermal equilibrium. On the other hand, some care is needed for the numerical calculation of the Laplace transform of $\langle \Delta x(t)^2\rangle$, as a particle trajectory of infinite duration would be in principle needed to determine the full dependence of $\langle \Delta x(\tau)^2\rangle$ on $\tau$ over the interval $0 \le \tau < \infty$ according to the definition of $\widetilde{\langle \Delta x(s)^2\rangle}$. Nevertheless, some simplifications can be carried out by taking into consideration the properties of $\langle \Delta x(\tau)^2 \rangle$ at thermal equilibrium inferred from~\eref{eq:msdx}, namely, $\langle \Delta x(0)^2 \rangle = 0$ and $\langle \Delta x(\tau \rightarrow \infty)^2 \rangle = \frac{2k_B T}{\kappa}$. Indeed, using the property $ \int_0^{\infty} d\tau \, e^{-s\tau}  \frac{d}{d \tau}f(\tau)  = s \tilde{f}(s) - f(0)$ of the Laplace transform of the derivative of a differentiable function, $f(\tau)$, such as 
$f(\tau) \equiv \langle \Delta x(\tau)^2 \rangle$, the Laplace transform of the mean-squared displacement for $s > 0$ can be expressed as
\begin{eqnarray}\label{eq:LapMSD}
    \widetilde{\langle \Delta x(s)^2\rangle} & = & \frac{1}{s} \langle \Delta x(0)^2 \rangle + \frac{1}{s} \int_0^{\infty} d\tau \,e^{-s\tau} \frac{d}{d\tau} \langle \Delta x(\tau)^2 \rangle, \nonumber\\
    & = & - \frac{1}{s}  \int_0^{\infty} d\tau \, \frac{d}{d\tau}\left( e^{-s\tau} \right) \langle \Delta x(\tau)^2 \rangle ,
\end{eqnarray}
where an integration by parts was performed in the second line of~\eref{eq:LapMSD}. We point out that in an actual optical tweezers experiment, the time lag $\tau$ is commonly discretized as $\tau_j = (j-1) \delta t$, with $j = 1, 2, \ldots$ and $\delta t$ the value of the time step set by the inverse of the acquisition frequency. Accordingly, the discretized version of~\eref{eq:LapMSD} reads
\begin{eqnarray}\label{eq:discLapMSD}
    \widetilde{\langle \Delta x(s)^2\rangle}& = & 
    \frac{1}{s} \sum_{j = 1}^{\infty} \frac{ \langle \Delta x_{j-1}^2 \rangle +  \langle \Delta x_j^2 \rangle}{2} \left( e^{-s \tau_{j-1}} - e^{-s \tau_j} \right), \nonumber\\
    & = & \frac{1}{s} \sum_{j = 1}^N  \frac{\langle \Delta x_{j-1}^2 \rangle +  \langle \Delta x_j^2 \rangle }{2} \left( e^{-s \tau_{j-1}} - e^{-s \tau_j} \right)  +  \frac{2k_B T}{s \kappa} e^{- s \tau_N},
\end{eqnarray}
where $\langle \Delta x_j^2\rangle$ represents the value of the mean-squared displacement computed at $\tau_j$, whereas $N$ is an integer such that $\langle \Delta x_N^2\rangle$ has already converged to the asymptotic value $\langle \Delta x(\tau \rightarrow \infty)^2 \rangle = \frac{2k_B T}{\kappa}$, which implies that, for all $j \ge N$, $\langle \Delta x_j^2\rangle = \frac{2k_B T}{\kappa}$. For the typical values of $\delta t$ and $\kappa$ in the experiments, $N \sim 10^4$. Therefore, the summation over the indices $j \ge N+1$ becomes
\begin{eqnarray}\label{eq:simp}
    \sum_{j=N+1}^{\infty} \frac{\langle \Delta x_{j-1}^2 \rangle +  \langle \Delta x_j^2 \rangle }{2} \left( e^{-s \tau_{j-1}} - e^{-s \tau_j} \right) & = & \frac{2k_B T}{\kappa} \sum_{j = N+1}^{\infty}\left[e^{-s\tau_N} - e^{-s\tau_{N+1}} \right. \nonumber\\
    & &  +e^{-s\tau_{N+1}} - e^{-s\tau_{N+2}}\nonumber\\
   && \left. +e^{-s\tau_{N+2}} - \ldots \right],\nonumber\\
   & = & \frac{2k_B T }{\kappa} e^{-s\tau_N},
\end{eqnarray}
where we have used the fact that all terms with $j \ge N+1$ cancel out in pairs on the right hand side of~\eref{eq:simp}, thereby simplifying to the single term corresponding to $j= N$, from which the second line of~\eref{eq:discLapMSD} was obtained. It should be noted that~\eref{eq:discLapMSD} only requires the knowledge of  the first $N$ discrete values of the mean-squared displacement in order to numerically compute the corresponding Laplace transform for $s > 0$. Moreover, since all the terms in the summation on the right-hand side of~\eref{eq:discLapMSD} are positive, this implies that $ \frac{2k_B T}{s \widetilde{\langle \Delta x(s)^2\rangle}}  < \kappa e^{s\tau_N} $ for any frequency $s >0$. This inequality limits the minimum frequency that can be reliably resolved in practice according to the generalized Stokes-Einstein relation~\eref{eq:GSER}. Indeed, it implies that $6\pi a s \tilde{\eta}(s) < \kappa \left( e^{s\tau_N } -1\right)$, which requires that $s \tau_N \gtrsim 1$ for a fixed value of $\kappa$ in order to resolve the low-frequency regime of $\tilde{\eta}(s)$ that is a monotonically decreasing function of $s$. Therefore, such a minimum frequency can be chosen as $s_{\mathrm{min}} = (N \delta t)^{-1}$, whereas the maximum frequency is simply set by the acquisition sampling rate of the experiment, i.e. $s_{\mathrm{max}} = \delta t^{-1}$.
Thus, \eref{eq:kap} and \eref{eq:discLapMSD} provide the basis for the calculation of the frequency-dependent viscosity of the surrounding fluid as well as the friction memory kernel experienced by the trapped particle over the frequency range $[s_{\mathrm{min}}, s_{\mathrm{max}}]$ from the experimental data for the stochastic time evolution of the particle position.


\begin{thebibliography}{99}

\bibitem{ciliberto2017} Ciliberto S 2017 {\it Phys. Rev. X} {\bf 7} 021051 

\bibitem{sekimoto1998} Sekimoto K 1998 {\it Prog Theor Phys Suppl.} {\bf 130} 17

\bibitem{seifert2012} Seifert U 2012 {\it Rep. Prog. Phys.} {\bf 75} 126001


\bibitem{jarzynski1997} Jarzynski C 1997 {\it Phys. Rev. Lett.} {\bf 78} 2690 

\bibitem{crooks1998} Crooks G E 1998 {\it J. Stat. Phys.} {\bf 90} 1481

\bibitem{hatano2001} Hatano T and Sasa S-i 2001 {\it Phys. Rev. Lett.} {\bf 86} 3463 

\bibitem{evans2002} Evans D J and Searles D J 2002 {\it Adv. Phys.} {\bf 51} 1529 

\bibitem{seifert2005} Seifert U 2005 {\it Phys. Rev. Lett.} {\bf 95} 040602 


\bibitem{harada2005} Hatano T and Sasa S-i 2005 {\it Phys. Rev. Lett.} {\bf 95} 130602 


\bibitem{chetrite2008} Chetrite R, Falkovich G and Gawedzki K 2008 {\it J. Stat. Mech.} P08005

\bibitem{baiesi2009} Baiesi M, Maes C and Wynants B
2009 {\it Phys. Rev. Lett.} {\bf 103} 010602 

\bibitem{seifert2010} Seifert U and Speck T 2010 {\it EPL} {\bf 89} 10007 

\bibitem{verley2011} Verley G, Ch\'etrite R and Lacoste D 2011 {\it J. Stat. Mech.} P10025

\bibitem{altaner2016} Altaner B, Polettini M and Esposito M 2016 {\it Phys. Rev. Lett.} {\bf 117} 180601 


\bibitem{wang2002} Wang G M, Sevick E M, Mittag E, Searles D J and Evans D J 2002 {\it Phys. Rev. Lett.} {\bf 89} 050601 


\bibitem{liphardt2002} Liphardt J, Dumont S, Smith S B, Tinoco Jr I and
Bustamante C 2002 {\it Science} {\bf 296} 1832

\bibitem{carberry2004} Carberry D M, Reid J C, Wang G M, Sevick E M, Searles D J and Evans D J 2004
{\it Phys. Rev. Lett.} {\bf 92}, 140601 


\bibitem{blickle2006} Blickle V, Speck T, Helden L, Seifert U and Bechinger C 2006 {\it Phys. Rev. Lett.} {\bf 96} 070603 


\bibitem{toyabe2007} Toyabe S, Jiang H-R, Nakamura T, Murayama Y and Sano M 2007 {\it Phys. Rev. E} {\bf 75} 011122 


\bibitem{gomezsolano2009} Gomez-Solano J R, Petrosyan A, Ciliberto S, Chetrite R and Gawedzki K 2009 {\it Phys. Rev. Lett.} {\bf 103} 040601 

\bibitem{gomezsolano2012}  Gomez-Solano J R, Petrosyan A and Ciliberto S 2012 {\it EPL} {\bf 98} 10007


\bibitem{ganguly2013} Ganguly C and Chaudhuri D 2013 {\it Phys. Rev. E} {\bf 88} 032102 

\bibitem{speck2016} Speck T 2016 {\it EPL} {\bf 114} 30006

\bibitem{dabelow2019} Dabelow L, Bo S and Eichhorn R 2019 {\it Phys. Rev. X} {\bf 9} 021009  

\bibitem{szamel2019} Szamel G 2019 {\it Phys. Rev. E} {\bf 100} 050603(R) 

\bibitem{cao2022} Cao Z, Su J, Jiang H and Hou Z 2022 {\it Phys. Fluids} {\bf 34} 053310 

\bibitem{hartich2021} Hartich D and Godec A 2021 {\it Phys. Rev. Lett.} {\bf 127} 080601 

\bibitem{khadem2022} Khadem S M J, Klages R and Klapp S H L 2022 {\it Phys. Rev. Research} {\bf 4} 043186 

\bibitem{fogedby2020} Fogedby H C 2020 {\it J. Stat. Mech.} 083208

\bibitem{imparato2007} Imparato A, Peliti L, Pesce G, Rusciano G and Sasso A 2007 {\it Phys. Rev. E} {\bf 76} 050101(R) 


\bibitem{chatterjee2010} Chatterjee D and Cherayil B J 2010 {\it Phys. Rev. E 82} {\bf 051104} 

\bibitem{chatterjee2011} Chatterjee D and Cherayil B J 2011 {\it J. Stat. Mech.} P03010


\bibitem{gomezsolano2011} Gomez-Solano J R, Petrosyan A and Ciliberto S 2011 {\it Phys. Rev. Lett.} {\bf 106} 200602 

\bibitem{crisanti2017} Crisanti A, Sarracino A and Zannetti M 2017 {\it Phys. Rev. E} {\bf 95} 052138 

\bibitem{denzler2018} Denzler T and Lutz E 2018 {\it Phys. Rev. E} {\bf 98} 052106 

\bibitem{chen2021} Chen J-F, Qiu T and Quan H-T 2021 {\it Entropy} {\bf 23} 1602

\bibitem{ciliberto_2013} Ciliberto S, Imparato A, Naert A and Tanase M 2013 {\it J. Stat. Mech.} P12014 

\bibitem{ciliberto2013} Ciliberto S, Imparato A, Naert A and Tanase M 2013 {\it Phys. Rev. Lett.} {\bf 110} 180601 

\bibitem{berut2016} B\'erut A, Imparato A, Petrosyan A and Ciliberto S 2016 {\it Phys. Rev. Lett.} {\bf 116} 068301 

\bibitem{saito2011} Saito K and Dhar A 2011 {\it Phys. Rev. E} {\bf 83} 041121 

\bibitem{kundu2011} Kundu A, Sabhapandit S and Dhar A 2011 {\it J. Stat. Mech.} P03007

\bibitem{fogedby2012} Fogedby H C and Imparato A 2012 {\it J. Stat. Mech.} P04005

\bibitem{dhar2015} Dhar A and Dandekar R 2015 {\it Physica A} {\bf 418} 49 

\bibitem{fogedby2009} Fogedby H C and Imparato A 2009 {\it J. Phys. A: Math. Theor.} {\bf 42} 475004

\bibitem{paraguassu_2021} Paraguass\'u P V and Morgado W A M 2021 {\it J. Stat. Mech.} 023205

\bibitem{paraguassu_2022} Paraguass\'u P V, Morgado W A M 2022 {\it Physica A} {\bf 588} 126576 


\bibitem{rosinberg2016} Rosinberg M L, Tarjus G and Munakata T 2016 {\it EPL} {\bf 113} 10007

\bibitem{salazar2016} Salazar D S P and Lira S A 2016 {\it J. Phys. A: Math. Theor.} {\bf 49} 465001

\bibitem{paraguassu2022} Paraguass\'u P V, Aquino R, Defaveri L and Morgado W A M 2022 {\it Phys. Rev. E} {\bf 106} 044106 

\bibitem{colmenares2022} Colmenares P J 2022 {\it Phys. Rev. E} {\bf 105} 044109 

\bibitem{paraguassu_2023} Paraguass\'u P V, Aquino R and Morgado W A M 2023 {\it Physica A} {\bf 615} 128568 

\bibitem{paraguassu2021}  Paraguass\'u P V and Morgado W A M 2021 {\it Eur. Phys. J. B} {\bf 94} 197


\bibitem{paraguassu2023} Paraguass\'u P V, Defaveri L and Morgado W A M 2023 {\it Eur. Phys. J. B} {\bf 96} 22 


\bibitem{gupta2021} Gupta D and Sivak D A 2021 {\it Phys. Rev. E} {\bf 104} 024605 

\bibitem{goswami2019} Goswami K 2019 {\it Phys. Rev. E} {\bf 99} 012112 

\bibitem{goswami2022} Goswami K 2022 {\it Phys. Rev. E} {\bf 105} 044123 

\bibitem{sarkar2023} Sarkar R, Santra I and Basu U 2023 {\it Phys. Rev. E} {\bf 107} 014123


\bibitem{larson1999} Larson R G 1999 {\it The Structure and Rheology of Complex Fluids} (New York: Oxford University Press)


\bibitem{zwanzig1973} Zwanzig R 1973 {\it Journal of Statistical Physics} {\bf 9} 215

\bibitem{chatterjee2009} Chatterjee D and Cherayil B J 2009 {\it Phys. Rev. E} {\bf 80} 011118

\bibitem{gomezsolano2015} Gomez-Solano J R and Bechinger C 2015 {\it New J. Phys.} {\bf 17} 103032 


\bibitem{roichman2021} Svetlizky I and Roichman Y 2021 {\it Phys. Rev. Lett.} {\bf 127} 038003

\bibitem{krishnamurthy2016} Krishnamurthy S, Ghosh S, Chatterji D, Ganapathy R and Sood A K 2016 {\it Nat. Phys.} {\bf 12} 1134


\bibitem{gomezsolano2021} Gomez-Solano J R 2021 {\it Front. Phys.} {\bf 9} 643333


\bibitem{guevara2023} Guevara-Valadez C A, Marathe R and Gomez-Solano J R 2023 {\it Physica A} {\it 609} 128342 


\bibitem{nalupurackal2023} Nalupurackal G et al 2023 {\it New J. Phys.} {\bf 25} 063001


\bibitem{gomezsolano2016} Gomez-Solano J R, Blokhuis A and Bechinger C 2016 {\it Phys. Rev. Lett.} {\bf 116} 138301 

\bibitem{narinder2018} Narinder N, Bechinger C and Gomez-Solano J R 2018 {\it Phys. Rev. Lett.} {\bf 121} 078003 

\bibitem{saad2019} Saad S and Natale G 2019 {\it Soft Matter} {\bf 15} 9909 


\bibitem{raman2023} Raman H, Das S, Sharma H, Singh K, Gupta S and Mangal R 2023 {\it ACS Phys. Chem Au} {\bf 3} 279


\bibitem{indei2012} Indei T, Schieber J D, Cordoba A and Pilyugina E 2012 {\it Phys. Rev. E} {\bf 85} 021504 

\bibitem{kim2005} Kim S and Karrila S J 2005 {\it Microhydrodynamics: Principles and Selected Applications} (New York: Dover Publications)


\bibitem{procopio2023} Procopio G and Giona M 2023 {\it Fluids} {\bf 8} 84

\bibitem{landau1959} Landau L D and Lifshitz E M 1959 {\it Fluid Mechanics} (London: Pergamon Press)

\bibitem{zwanzig1970} Zwanzig R and Bixon M 1970 {\it Phys. Rev. A} {\bf 2} 2005

\bibitem{kubo1991} Kubo R, Toda M and Hashitsume N 1991 {\it Statistical Physics II: Nonequilibrium Statistical Mechanics} 2nd edition (Berlin: Springer-Verlag)

\bibitem{adelman1977} Adelman S A and Garrison B J 1977 {\it Molec. Phys.} {\bf 33} 1671 

\bibitem{okuyama1986} Okuyama S and Oxtoby D W 1986 {\it J. Chem. Phys.} {\bf 84} 5824 

\bibitem{roux1999} Roux B and Simonson T 1999 {\it Biophys. Chem.} {\bf 78} 1 

\bibitem{gelin2009} Gelin M F and Thoss M 2009 {\it Phys. Rev. E} {\bf 79} 051121 

\bibitem{seifert2016} Seifert U 2016 {\it Phys. Rev. Lett.} {\bf 116} 020601  

\bibitem{ding2022} Ding M, Liu F and Xing X 2022 {\it Phys. Rev. Research} {\bf 4} 013015

\bibitem{talkner2016} Talkner P and H\"anggi P 2016
{\it Phys. Rev. E} {\bf 94} 022143 

\bibitem{talkner2020} Talkner P and H\"anggi P 2020
{\it Rev. Mod. Phys.} {\bf 92} 041002 

\bibitem{gradshteyn1980} Gradshteyn I S and Ryzhik I M 1980 {\it Table of Integrals, Series
and Products} (San Diego: Academic Press)

\bibitem{lam2019} Lam C N, Do C, Wang Y, Huang G-R and Chen W-R 2019 {\it Phys.Chem.Chem.Phys.} {\bf 21} 18346

\bibitem{ebagninin2009} Ebagninin K W, Benchabane A and Bekkour K 2009 {\it J. Colloid Interface Sci.} {\bf 336} 360

\bibitem{yasuda1981} Yasuda K, Armstrong R C and Cohen R E 1981 {\it Rheol. Acta} {\bf 20} 163

\bibitem{cross1979} Cross M M 1979 {\it Rheol. Acta} {\bf 18} 609


\bibitem{talbot1979} Talbot A 1979 {\it IMA J. Appl. Math.} {\bf 23} 97


\bibitem{strasberg2017} Strasberg P and Esposito M 2017 {\it Phys. Rev. E} {\bf 95} 062101 

\bibitem{aurell2018} Aurell E 2018 {\it Phys. Rev. E} {\bf 97} 042112 

\bibitem{cockrell2022} Cockrell C and Ford I J 2022 {\it Phys. Rev. E} {\bf 105} 064124 

\bibitem{venturelli2022} Venturelli D, Ferraro F and Gambassi A 2022 {\it Phys. Rev. E} {\bf 105} 054125
 
\bibitem{carberry2007}  Carberry D M, Baker M A B, Wang G M, Sevick E M and Evans D J 2007 {\it J. Opt. A: Pure Appl. Opt.} {\bf 9} S204 

\bibitem{huang2021} Huang D, Lu S, Shi X, Goree J and Feng Y 2021 {\it Phys. Rev. E} {\bf 104} 035207 

\bibitem{narinder2021} Narinder N, Paul S and Bechinger C 2021 {\it Phys. Rev. E} {\bf 104} 034605

\bibitem{ferrer2021} Ferrer B R, Gomez-Solano J R and Arzola A V 2021 {\it Phys. Rev. Lett.} {\bf 126} 108001


\bibitem{kundu2021} Kundu A, Dey R, Paul S and Banerjee A 2021 {\it Phys. Rev. Fluids} {\bf 6} 123301

\bibitem{das2023} Das B, Paul B, Manikandan S K and Banerjee A 2023 Enhanced directionality of active processes in a viscoelastic bath arXiv:2302.01996 


\end{thebibliography}
\end{document}